\documentclass[twocolumn,journal]{IEEEtran}
\usepackage[usenames,dvipsnames]{color}
\usepackage{subfigure}
\usepackage{graphicx}
\usepackage{amsmath}
\usepackage{color}
\usepackage{multicol}
\usepackage{amssymb}
\usepackage{amsthm}

\usepackage{amsfonts}%
\usepackage{verbatim}%
\usepackage{enumerate}%
\usepackage{psfrag}
\usepackage{epsfig}
\usepackage{multicol}
\usepackage{setspace}
\usepackage{dsfont}
\usepackage{balance}

\usepackage{algorithm}
\usepackage{algorithmic}
\usepackage{cite}
\usepackage{float}

\ifCLASSINFOpdf
\else
\fi
\begin{document}
\title{Optimal Random Access and Random Spectrum Sensing for an Energy Harvesting Cognitive Radio with and without Primary Feedback Leveraging}
\author{Ahmed~El~Shafie,~\IEEEmembership{Member,~IEEE,}
\thanks{Part of this work has been presented in the 8th International Conference on Wireless and Mobile Computing, Networking and Communications (WiMob), 2012 \cite{wimob}
}
\thanks{A. El Shafie is with Wireless Intelligent Networks Center (WINC), Nile University, Giza, Egypt. e-mail: ahmed.salahelshafie@gmail.com}}
%\markboth{El~Shafie: \footnotesize {Optimal Random Access and Random Spectrum Sensing for an Energy Harvesting Cognitive Radio with and without Primary Feedback Leveraging}}
\date{}
\maketitle

\begin{abstract}
We consider a secondary user (SU) with energy harvesting capability. We design access schemes for the SU which incorporate random spectrum sensing and random access, and which make use of the primary automatic repeat request (ARQ) feedback. We study two problem-formulations. In the first problem-formulation, we characterize the stability region of the proposed schemes. The sensing and access probabilities are obtained such that the secondary throughput is maximized under the constraints that both the primary and secondary queues are stable. Whereas in the second problem-formulation, the sensing and access probabilities are obtained such that the secondary throughput is maximized under the stability of the primary queue and that the primary queueing delay is kept lower than a specified value needed to guarantee a certain quality of service (QoS) for the primary user (PU). We consider spectrum sensing errors and assume multipacket reception (MPR) capabilities. Numerical results show the enhanced performance of our proposed systems.
\end{abstract}

\section{Introduction}
Cognitive radio technology provides an efficient means of utilizing the radio spectrum \cite{zhao2007survey}. The basic idea is to allow secondary users to access the spectrum while providing certain guaranteed quality of service (QoS) performance measures for the primary users (PUs). The secondary user (SU) is a battery-powered device in many practical situations and its operation, which involves spectrum sensing and access, is accompanied by energy consumption. Consequently an energy-constrained SU must optimize its sensing and access decisions to efficiently utilize the energy at its disposal. An emerging technology for energy-constrained terminals is energy harvesting which allows the terminal to gather energy from its environment. An overview of the different energy harvesting technologies is provided in \cite{survey} and the references therein.

Data transmission by an energy harvester with a rechargeable battery has got a lot of attention recently  \cite{lei2009generic,sharma2010optimal,ho2010optimal,yang2010transmission,yang2010optimal,tutuncuoglu2010optimum,pappas2011optimal,krikidis2012stability,Sultan,ourletter}. The optimal online policy for controlling admissions into the data buffer
is derived in \cite{lei2009generic} using a dynamic programming framework. In \cite{sharma2010optimal},
energy management policies which stabilize the data queue
are proposed for single-user communication and some delay-optimal properties are derived. Throughput optimal
energy allocation is investigated in \cite{ho2010optimal} for energy harvesting systems in
a time-constrained slotted setting. In \cite{yang2010transmission,yang2010optimal}, minimization
of the transmission completion time is considered in an
energy harvesting system and the optimal solution is obtained
using a geometric framework. In \cite{tutuncuoglu2010optimum}, energy harvesting transmitters with
batteries of finite energy storage capacity are considered and
the problem of throughput maximization by a deadline is
solved for a static channel.

The authors of \cite{pappas2011optimal} consider the scenario in which a set of
nodes shares a common channel. The PU has a rechargeable
battery and the SU is plugged to a reliable power supply. They obtain the maximum stable throughput region which describes the maximum arrival rates that maintain the stability of the network queues. In \cite{krikidis2012stability}, the authors investigate the effects of network layer
cooperation in a wireless three-node network with energy harvesting
nodes and bursty traffic. In \cite{Sultan}, Sultan investigated the optimal cognitive sensing and access policies for an SU with an energy queue. The analysis is based on Markov-decision process (MDP). In \cite{ourletter}, the authors investigated the maximum stable throughput of a backlogged secondary terminal with energy harvesting capability. The SU randomly accesses the channel at the beginning of the time slot without employing any channel sensing. The secondary terminal can leverage the availability of primary feedback and exploit the multipacket reception (MPR) capability of the receivers to enhance its throughput.

In this paper, we develop spectrum sensing and transmission methods for an energy harvesting SU. We leverage the primary automatic repeat request (ARQ) feedback for secondary access. Due to the broadcast nature of the wireless channel, this feedback can be overheard and utilized by the secondary node assuming that it is unencrypted. The proposed protocols can alleviate the negative impact of channel sensing because the secondary access is on basis of the sensed primary state as well as the overheard primary feedback. The problem with depending on spectrum sensing only is that sensing does not inform the secondary terminal about its impact on the primary receiver. This issue has induced interest in utilizing the feedback from the primary receiver to the primary transmitter to optimize the secondary transmission
strategies. For instance, in \cite{eswaran2007bits}, the SU observes the ARQ feedback from the primary
receiver as it reflects the PU's achieved packet rate. The SU's objective is to maximize its throughput while guaranteeing a certain packet rate for the PU. In \cite{lapiccirella2010cognitive}, the authors use a partially observable Markov decision process (POMDP) to optimize the
secondary action on the basis of the spectrum sensing outcome and primary ARQ feedback. Secondary power control based on primary feedback is investigated in \cite{huang2010distributed}. In \cite{levorato2009cognitive} and \cite{levorato2012cognitive}, the optimal transmission policy for the SU when the PU adopts a retransmission based error control scheme is investigated. The policy of the SU determines how often it transmits according to the retransmission state of the packet being served by the PU.

The contributions of this paper can be summarized as follows.
\begin{itemize}
\item We investigate the case of an SU equipped with an energy harvesting mechanism and a rechargeable battery.
\item We propose a novel access and sensing schemes where the SU possibly senses the channel for a certain fraction of the time slot duration and accesses the channel with some access probability that depends on the sensing outcome. The SU may access the channel probabilistically without sensing in order to utilize the whole slot duration for transmission. Furthermore, it leverages the primary feedback signals.
    \item Instead of the collision channel model, we assume a generalized channel model in which the receiving nodes have MPR capability.
    \item We propose two problem-formulations. In the first problem-formulation, we characterize the stability region of the proposed schemes. The sensing and access probabilities are obtained such that the secondary throughput is maximized under the constraints that both the primary and secondary queues are stable. Whereas in the second problem-formulation, we include a constraint on the primary queueing delay to the optimization problem for delay-aware PUs. The sensing and access probabilities are obtained such that the secondary throughput is maximized under the stability of the primary queue and that the primary queueing delay is kept lower than a specified value needed to guarantee a certain QoS for the PU.
%        \item For the stability region of the proposed protocols, we provide two dominant systems for each protocol, which consider as an inner bound of the stability region
        \item We compare our systems with the conventional access system in which the SU senses the channel and accesses unconditionally if the PU is sensed to be inactive. The numerical results show the gains of our proposed systems in terms of the secondary throughput.
\end{itemize}
\begin{figure}
  \includegraphics[width=1.05\columnwidth]{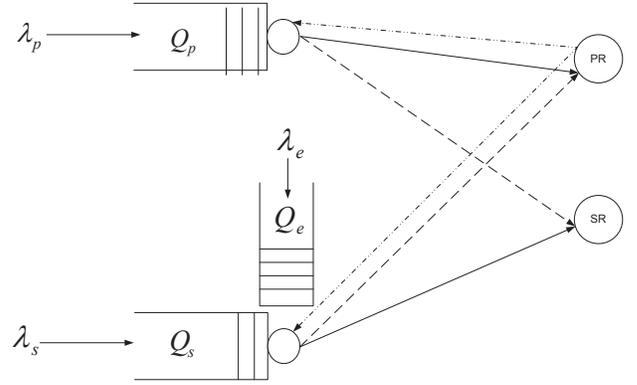}\\
  \caption{Primary and secondary queues and links. The PU has data queue $Q_p$, whereas the secondary terminal has data queue $Q_s$ and energy queue $Q_e$. There is a feedback channel between the primary receiver (PR) and the PU to acknowledge the reception of data packets. This feedback channel is overheard by the secondary transmitter. Both the PR and the secondary receiver (SR) may suffer interference from the other link. }\label{fig1}
\end{figure}

The rest of the paper is organized as follows. In the next section, we discuss the system model adopted in this paper. The secondary access without incorporating the primary feedback is investigated in Section \ref{sec2}. In Section \ref{sec3}, we discuss the feedback-based scheme. The case of delay-aware PUs is investigated in Section \ref{delayaware}. We provide numerical results and conclusions in Section \ref{numerical}.
% and conclude the paper in Section \ref{secfinal}.

%%%%%%%%%%%%%%
%%%%%%%%%%%%%%%%%
%%%%%%%%%%%%%%%%%
%%%%%%%%%%%%%%
\section{System Model}
\label{systemmodel}
We consider the system model shown in Fig. \ref{fig1}. The model consists of one PU and
one SU. The channel is slotted in time and
a slot duration equals the packet transmission time. The PU and the SU have infinite buffer queues, $Q_p$ and $Q_s$, respectively, to store fixed-length data packets. If a terminal transmits during a time slot, it sends exactly one packet to its receiver. The
arrivals at $Q_p$ and $Q_s$ are independent and identically
distributed (i.i.d.) Bernoulli random variables from slot to slot with means $\lambda_p$ and $\lambda_s$, respectively.

The SU has an additional energy queue, $Q_e$, to store harvested energy from the environment. The arrival at the energy queue is also Bernoulli with mean $\lambda_e$ and is independent from arrivals at the other queues. The Bernoulli model is simple, but it captures the random availability of ambient energy sources. More importantly, in the analysis of discrete-time queues, Bernoulli arrivals see time averages (BASTA). This is the BASTA property equivalent to
the Poisson arrivals see time averages (PASTA) property in continuous-time systems \cite{basta1}. It is assumed that the transmission of one data packet consumes one packet of energy.

We adopt a late arrival model as in \cite{sadek,pappas2011optimal,krikidis2012stability,ourletter} where an arrived packet at a certain time slot cannot be served at the arriving slot even if the queue is empty. Denote by $\mathcal{V}^t$ the number of arrivals to queue $Q$
at time slot $t$, and $\mathcal{Z}^t$ the number of departures from queue $Q$ at time slot $t$. The queue length evolves according to the following form:
\begin{equation}
    Q^{t+1}=(Q^t-\mathcal{Z}^t)^+ +\mathcal{V}^t
\end{equation}
where $(z)^+$ denotes $\max(z,0)$.

Adequate system operation requires that all the queues are stable. We employ the standard definition of stability for queue as in \cite{sadek,stabN}, that is, a queue is stable if and only if its probability of being empty does not vanish as time progresses. Precisely,
    $ \lim_{t \rightarrow \infty  }{\rm Pr}\{Q^t=\tilde 0\}>0$.
If the arrival and service processes are strictly stationary, then we can apply Loynes theorem to check for stability conditions \cite{loynes1962stability}. This theorem states that if the arrival process and the service process of a queue are strictly stationary processes, and the average service rate is greater than the average arrival rate of the queue, then the queue is stable. If the average service rate is lower than the average arrival rate, the queue is unstable \cite{sadek}.

Instead of the collision channel model where simultaneous transmission by different terminals leads to sure packet loss, we assume that the receivers have MPR capability as in \cite{ghez1988stability,ghez1989optimal,naware2005stability}. This means that transmitted data packets can survive the interference caused by concurrent
transmissions if the received signal to interference
and noise ratio ({\rm SINR}) exceeds the threshold required for successful decoding at the receiver. With MPR capability, the SU may use the channel simultaneously with the PU.

\section{Secondary Access Without Employing Primary Feedback}\label{sec2}
The first proposed system is denoted by $\Phi_{\rm NF}$. Under this protocol, the PU accesses the channel whenever
it has a packet to send. The secondary transmitter, given that it has energy, senses the channel or possibly transmits the packet at the head of its queue immediately at the beginning of the time slot without sensing the channel. We explain below why direct transmission can be beneficial for system performance.

The SU operation can be summarized as follows.
\begin{itemize}
  \item If the secondary terminal's energy and data queues are not empty, it senses the channel with probability $p_s$
   from the beginning of the time slot for a duration of $\tau$ seconds to detect the possible activity of the PU. If the slot duration is $T$, $\tau < T$.
    \item If the channel is sensed to be free, the secondary transmitter accesses the channel with probability $p_f$. If the PU is detected to be active, it accesses the channel with probability $p_b$.
    \item If at the beginning of the time slot the SU decides not to sense the spectrum (which happens with probability $1-p_s$), it immediately decides whether to transmit with probability $p_t$ or to remain idle for the rest of the time slot with probability $\overline{p_t}=1-p_t$.\footnote{Throughout the paper $\overline{\mathcal{X}}\!=\!1\!-\!\mathcal{X}$.}
\end{itemize}
\noindent This means that the transmission duration is $T$ seconds if the SU accesses the channel without spectrum sensing and $T-\tau$ seconds if transmission is preceded by a sensing phase. We assume that the energy consumed in spectrum sensing is negligible, whereas data transmission dissipates exactly one unit of energy from $Q_e$.

\begin{table}
\renewcommand{\arraystretch}{1}
\begin{center}
\begin{tabular}{ |c |l||  }
    \hline\hline
    $\tau$ & {\footnotesize Sensing duration} \\[5pt]\hline
  $T$ & {\footnotesize Slot duration} \\[5pt]\hline
  $p_s$ & {\footnotesize Probability of sensing the channel} \\[5pt]\hline
   $P_{\rm MD}$ & {\footnotesize Misdetection probability} \\[5pt]\hline
    $P_{\rm FA}$ & {\footnotesize False alarm probability} \\[5pt]\hline
  $p_t$ & {\footnotesize Probability of direct channel access if the channel is } \\
   & {\footnotesize not sensed} \\[5pt]\hline
  $p_f$ & {\footnotesize Probability of channel access if the channel is sensed} \\
   & {\footnotesize to be free} \\[5pt]\hline
    $p_b$ & {\footnotesize Probability of channel access if the channel is sensed} \\
   & {\footnotesize to be busy} \\[5pt]\hline & \\
 $\overline{P}_{p}$ & {\footnotesize Probability of successful primary transmission to the } \\
   & {\footnotesize primary receiver if the secondary terminal is silent}\\[5pt]\hline &\\
   $\overline{P_{p}^{\left({\rm c}\right)}}$ & {\footnotesize Probability of successful primary transmission to the } \\
   & {\footnotesize primary receiver with concurrent secondary transmission}\\[5pt]\hline & \\
 $\overline{P}_{0s}$ & {\footnotesize Probability of successful secondary transmission if the PU} \\
   & {\footnotesize is silent and transmission occurs over $T$ seconds}\\[5pt]\hline &\\
   $\overline{P}_{1s}$ & {\footnotesize Probability of successful secondary transmission if the PU} \\
   & {\footnotesize is silent and transmission occurs over $T-\tau$ seconds}\\[5pt]\hline &\\
   $\overline{P_{0s}^{\left({\rm c}\right)}}$ & {\footnotesize Probability of successful secondary transmission if the PU} \\
   & {\footnotesize is active and transmission occurs over $T$ seconds}\\[5pt]\hline & \\
   $\overline{P_{1s}^{\left({\rm c}\right)}}$ & {\footnotesize Probability of successful secondary transmission if the PU} \\
   & {\footnotesize is active and transmission occurs over $T-\tau$ seconds}\\[5pt]\hline
\end{tabular}
\caption{List of symbols involved in the queues' mean service rates.}
\label{table1}
\end{center}
\end{table}

We study now secondary access in detail to obtain the mean service rates of queues $Q_s$, $Q_e$ and $Q_p$. The meaning of the various relevant symbols are provided in Table \ref{table1}. For the secondary terminal to be served, its energy queue must be nonempty. If the SU does not sense the channel, which happens with probability $1-p_s$, it transmits with probability $p_t$. If the PU's queue is empty and, hence, the PU is inactive, secondary transmission is successful with probability $\overline{P}_{0s}$, whose expression as a function of the secondary link parameters, transmission time $T$, and the data packet size is provided in Appendix A. If $Q_p \neq 0$, secondary transmission is successful with probability $\overline{P_{0s}^{\left({\rm c}\right)}}$ (see Appendix A).
If the SU decides to sense the channel, there are four possibilities depending on the sensing outcome and the state of the primary queue. If the PU is sensed to be free, secondary transmission takes place with probability $p_f$. This takes place with probability $1-P_{\rm FA}$ if the PU is actually silent. In this case, the probability of successful secondary transmission is $\overline{P}_{1s}$, which is lower than $\overline{P}_{0s}$ (for proof, see \cite{wimob}). On the other hand, if the PU is on, the probability of detecting the channel to be free is $P_{\rm MD}$ and the probability of successful secondary transmission is $\overline{P_{1s}^{\left({\rm c}\right)}}$. If the channel is sensed to be busy, the secondary terminal transmits with probability $p_b$. Sensing the PU to be active occurs with probability $P_{\rm FA}$ if the PU is actually inactive, or with probability $1\!-\!P_{\rm MD}$ if the PU is actively transmitting. The probability of successful secondary transmission is $\overline{P}_{1s}$ when the PU is silent and $\overline{P_{1s}^{\left({\rm c}\right)}}$ when the PU is active. Given these possibilities, we can write the following expression for the mean secondary service rate.
\begin{equation}
\begin{split}
\label{dodo}
\mu_s&=\left(1-p_s\right)p_t{\rm Pr}\{Q_p =0,Q_e \ne 0\} {\overline{P}}_{0s}
\\&+\left(1-p_s\right)p_t{\rm Pr}\{Q_p \ne 0,Q_e \ne 0\} \overline{P^{\left(\rm c\right)}_{0s}}
\\& +p_s p_f {\rm Pr}\{Q_p =0,Q_e \ne 0\} \left(1-P_{\rm FA}\right){\overline{P}}_{1s}
\\&+p_s p_f {\rm Pr}\{Q_p \ne0,Q_e \ne 0\} P_{\rm MD} \overline{P_{1s}^{\left({\rm c}\right)}}
\\&+p_s p_b {\rm Pr}\{Q_p =0,Q_e \ne 0\}P_{\rm FA} {\overline{P}}_{1s}
\\&+p_s p_b  {\rm Pr}\{Q_p \ne0,Q_e \ne 0\} \left(1-P_{\rm MD}\right) \overline{P_{1s}^{\left({\rm c}\right)}}
\end{split}
\end{equation}

 Based on the above analysis, it can be shown that the mean service rate of the energy queue is
\begin{equation}
\begin{split}
\mu_e&=\left(1-p_s \right)p_t{\rm Pr}\{Q_s \ne 0\}\\& +p_s p_f \bigg(P_{\rm MD} {\rm Pr}\{Q_s \ne 0,Q_p\ne 0\}\\& \,\,\,\,\,\,\,\,\,\,\,\,\,\,\,\,\ +(1-P_{\rm FA}){\rm Pr}\{Q_s \ne 0,Q_p=0\}\bigg) \\& +p_s p_b \bigg(P_{\rm FA} {\rm Pr}\{Q_s \ne 0,Q_p=0\}\\& \,\,\,\,\,\,\,\,\,\,\,\,\,\,\,\,\ +(1-P_{\rm MD}){\rm Pr}\{Q_s \ne 0,Q_p\ne 0\}\bigg)
\end{split}
\end{equation}

A packet from the primary queue can be served in either one of the following events. If the SU is silent because either of its data queue or energy queue is empty, the primary transmission is successful with probability $\overline{P}_{p}$. If both secondary queues are nonempty, secondary operation proceeds as explained above. In all cases, if the SU does not access the channel, the probability of successful primary transmission is $\overline{P}_{p}$, else it is $\overline{P_{p}^{\left({\rm c}\right)}}$.\footnote{We assume that the access delay of the SU does not affect the primary outage probability. This is valid as far as $(1-\frac{\tau}{T})e\approx e$, which is true here as $\tau\ll T$. For details, see Appendix A.} Therefore,
\begin{equation}
\begin{split}
\mu_p&=\bigg(1-{\rm Pr}\{Q_s \ne0,Q_e \ne 0\}\bigg)\overline{P}_{p}
\\&+{\rm Pr}\{Q_s \ne0,Q_e \ne 0\}\\&
\times \Bigg[\left(1-p_s\right)\bigg(p_t \overline{P_{p}^{\left({\rm c}\right)}}+\left(1-p_t\right)\overline{P}_{p}\bigg)
\\&\,\,\,\,\,\,\,\,\,\,\,\,\,\ +p_s P_{\rm MD}\bigg(p_f \overline{P_{p}^{\left({\rm c}\right)}}+\left(1-p_f\right)\overline{P}_{p}\bigg)
\\&\,\,\,\,\,\,\,\,\,\,\,\,\,\  +p_s \left(1-P_{\rm MD}\right)\bigg(p_b \overline{P_{p}^{\left({\rm c}\right)}}+\left(1-p_b\right)\overline{P}_{p}\bigg)\Bigg]
\label{mute}
\end{split}
\end{equation}
The maximum primary throughput is $\overline{P}_{p}$, i.e., $\mu_p\le \overline{P}_{p}$, which occurs when the PU operates alone, i.e., when the SU is always inactive.

Following are some important {\bf remarks} on the proposed access and sensing scheme. Firstly, the proposed access and sensing scheme can mitigate the negative impact of sensing errors. Specifically, the SU under the proposed protocol randomly accesses the channel if the PU is either sensed to be active or inactive. Hence, the false alarm probability and the misdetection probability are controllable using the spectrum access probabilities. These access probabilities can take any value between zero and one. Hence, the SU can mitigate the impact of the sensing errors via adjusting the values of the access probabilities. Accordingly, this would enhance the secondary throughput and prevent the violation of the PU's QoS.

Secondly, when the MPR capabilities of the receivers are strong (which means $\overline{P_{p}^{\left({\rm c}\right)}}\approx \overline{P}_{p}$ and  $\overline{P}_{is}^{\left({\rm c}\right)}\approx \overline{P}_{is}$ for $i=0,1$), the SU does not need to sense the channel at all, i.e., $p_s=0$. This is due to the fact that the SU does not need to employ channel sensing as it can transmit each time slot simultaneously with the PU without violating the primary QoS because the receivers can decode packets under interference with a probability almost equal to the decoding probability when nodes transmit alone.

 As in \cite{wimob,ourletter,pappas2011optimal}, we assume that the energy queue is modeled as M/D/1 queue with mean service and arrival rates $\mu_e=1$ and $\lambda_e$, respectively. Hence, the probability that the secondary energy queue being nonempty is $\lambda_e/\mu_e=\lambda_e$ \cite{kleinrock1975queueing}. Based on this assumption, the energy queue in the approximated system empties faster
than in the actual system, thereby lowering the probability that
the queue is nonempty. This reduces the secondary throughput
by increasing the probability that the secondary node does not
have energy. Therefore, our approximation result in a lower bound on the secondary service rate or
throughput \cite{wimob,ourletter}. We denote the system under the approximation of one packet consumption from the energy queue each time slot as $\mathcal{S}$.

 Since the queues in the approximated system, $\mathcal{S}$, are interacting with each other, we resort to the concept of the dominant system to obtain the stability region of system $\mathcal{S}$. The dominant system approach is first introduced in \cite{rao1988stability}. The basic idea is that we construct
an appropriate dominant system, which is a modification of
system $\mathcal{S}$ with the queues decoupled, hence we can compute the departure processes of all queues. The modified system ensures that the queue sizes in the
dominant system are, at all times, at least as large as those
of system $\mathcal{S}$ provided that the queues in both systems have the same initial sizes. Thus, the stability region of the new
system is an {\bf inner bound} of system $\mathcal{S}$. At the boundary points of the stability region, both the dominant system and system $\mathcal{S}$ coincide. This is the essence of the indistinguishability argument presented in many papers such as \cite{rao1988stability,sadek,pappas2011optimal,naware2005stability}. Next, we construct two dominant systems and the stability region of approximated system $\mathcal{S}$ is the union of the stability region of the dominant systems. We would like to emphasize here that system $\mathcal{S}$ is an {\bf inner bound} on the original system, $\Phi_{\rm NF}$.
\begin{figure}
  \includegraphics[width=1\columnwidth]{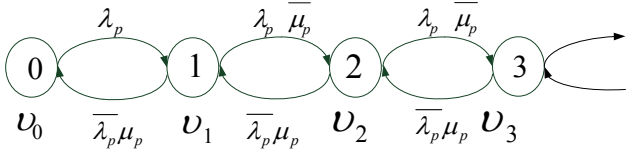}\\
  \caption{Markov chain of the PU under dominant system $\mathcal{S}_1$. State self-transitions are omitted for visual clarity. Probabilities $\overline{\mu_p}=1-\mu_p$ and $\overline{\lambda_p}=1-\lambda_p$.}\label{fig22}
\end{figure}

\subsection{First Dominant System}
In the first dominant system queue, denoted by $\mathcal{S}_1$, $Q_s$ transmits dummy packets when it is empty and the PU behaves as it would in the original system. Under this dominant system, we have ${\rm Pr}\{Q_s=0\}=0$. Substituting by ${\rm Pr}\{Q_s=0\}=0$ into (\ref{mute}), the average primary service rate after some simplifications can be given by
\begin{equation}
\begin{split}
\mu_p&=\overline{P}_{p}\!-\!\lambda_e\Delta_p\Bigg(\overline{p_s}p_t\!+\!p_s P_{\rm MD}p_f \!+\!p_s \overline{P_{\rm MD}}p_b\Bigg)
\label{mup}
\end{split}
\end{equation}
where $\Delta_p\!=\!\overline{P}_{p}\!-\! \overline{P_{p}^{\left({\rm c}\right)}}\!\ge\!0$. The Markov chain modeling the primary queue under this dominant system is provided in Fig. \ref{fig22}. Solving the state balance equations, it is straightforward to show that the probability that the primary queue has $k$ packets is
\begin{equation}
\nu_k=\nu_{0}\frac{1}{\overline{\mu_p}}\Bigg[\frac{\lambda_p \overline{\mu_p}}{\overline{\lambda_p }\mu_p}\Bigg]^{k}
\label{nuk}
\end{equation}
\noindent where $\overline{\lambda_p}=1-\lambda_p$ and $\overline{\mu_p}=1-\mu_p$. Using the condition $\sum_{k=0}^{\infty}\nu_k=1$,
\begin{equation}
\nu_{0}=1-\frac{\lambda_p}{\mu_p}
\end{equation}
\noindent For the sum $\sum_{k=0}^{\infty}\nu_k$ to exist, we should have $\lambda_p < \mu_p$. This is equivalent to Loynes' theorem. Since ${\rm Pr}\{Q_p=0\}=1-\frac{\lambda_p}{\mu_p}$,
\begin{equation}
\begin{split}
\mu_s&\!=\!\lambda_e \Bigg[\big(1\!-\!\frac{\lambda_p}{\mu_p}\big) \Bigg(\overline{p_s}p_t {\overline{P}}_{0s}\!+\!p_s p_b P_{\rm FA} {\overline{P}}_{1s}\!+\!p_s p_f  \overline{P_{\rm FA}}\ {\overline{P}}_{1s} \Bigg)
\\&\,\,\,\,\,\,\,\,\,\,\ +\!\frac{\lambda_p}{\mu_p}\Bigg(\overline{p_s}p_t \overline{P^{\left(\rm c\right)}_{0s}} +p_s p_f  P_{\rm MD} \overline{P_{1s}^{\left({\rm c}\right)}}\!+\!p_s p_b   \overline{P_{\rm MD}}\ \overline{P_{1s}^{\left({\rm c}\right)}} \Bigg)\Bigg]
\end{split}
\label{mus}
\end{equation}
Let \begin{equation}
\mu_{p}=\overline{P_p}+\mathcal{D}^\dagger \mathcal{P}
\label{mupvec}
\end{equation}
 where $(\bullet)^\dagger$ denotes vector transposition, $\mathcal{P}\!=\![p_t, p_b,p_f]^\dagger$ and $ \mathcal{D}\!=\!-\Delta_p \lambda_e\Big[\overline{p_s}, p_s \overline{P_{\rm MD}},p_s P_{\rm MD}\Big]^\dagger\!$.
 Substituting from (\ref{mupvec}) into (\ref{mus}),
\begin{equation}
\begin{split}
\mu_s&=\bigg(1\!-\!\frac{\lambda_p}{\overline{P}_{p}\!+\!\mathcal{D}^\dagger \mathcal{P}}\bigg) \mathcal{A}^\dagger \mathcal{P} +\frac{\lambda_p}{\overline{P}_{p}\!+\!\mathcal{D}^\dagger \mathcal{P}}\mathcal{G}^\dagger \mathcal{P}
\label{musxxfg}
\end{split}
\end{equation}
where $\mathcal{G}\!=\!\Big[\overline{p_s}p_t \overline{P^{\left(\rm c\right)}_{0s}},\!p_s    \overline{P_{\rm MD}}\ \overline{P_{1s}^{\left({\rm c}\right)}},p_s  P_{\rm MD} \overline{P_{1s}^{\left({\rm c}\right)}}\Big]$ and $\mathcal{A}\!=\!\Big[\overline{p}_s {\overline{P}}_{0s},p_s  P_{\rm FA} {\overline{P}}_{1s},p_s  \overline{P_{\rm FA}}~\overline{P}_{1s} \Big]^\dagger$.
After some mathematical manipulations, we get
\begin{equation}
\begin{split}
\mu_s&=\frac{(\overline{P}_{p}\!-\!\lambda_p)\mathcal{A}^\dagger \mathcal{P}\!+\! \mathcal{P}^\dagger\mathcal{D}\mathcal{A}^\dagger \mathcal{P}  \!+\!\lambda_p\mathcal{G}^\dagger \mathcal{P}}{\overline{P}_{p}\!+\!\mathcal{D}^\dagger \mathcal{P}}
\label{musxxfg2}
\end{split}
\end{equation}
The portion of the stability region based on the first dominant system is characterized by the closure of the rate pairs $(\lambda_p,\lambda_s)$. One method to obtain this closure is to solve a constrained optimization problem such that $\lambda_s$ is maximized for each $\lambda_p$ under the stability of the primary and the secondary queues. The optimization problem is given by
\begin{equation}
\begin{split}
& \underset{p_s,\mathcal{P}=[p_t,p_b,p_f]^\dagger}{\max.} \,\, \mu_s, \,\ {\rm s.t.} \,\,\,\,\  0 \le p_t,p_s,p_f,p_b \le 1,\,\ \lambda_p \le \mu_p
\label{opt1ofS}
\end{split}
\end{equation}
 For a fixed $p_s$, the optimization problem (\ref{opt1ofS}) can be shown to be a quasiconcave program over $\mathcal{P}$. We need to show that the objective function is quasiconcave over convex set and under convex constraints. From (\ref{mupvec}), $\mu_p$ is affine and hence convex over $\mathcal{P}$ for a fixed $p_s$. The Hessian of the numerator of $\mu_s$ is given by $H\!=\!\mathcal{A}\mathcal{D}^\dagger\!+\!\mathcal{D}\mathcal{A}^\dagger\!$.  Let $y$ be an arbitrary $3 \times 1$ vector. The matrix $H$ is negative semidefinite if $y^\dagger H y\!\le\!0$.
Since the matrices $\mathcal{A}\mathcal{D}^\dagger$ and $\mathcal{D}\mathcal{A}^\dagger$ are generated using a linear combination of a single vector, the rank of each is $1$ and therefore each of them has at least two zero eigenvalues. The trace of each is negative and equal to $\Lambda\!=\!-\Delta_p \lambda_e (\overline{p}^2_s {\overline{P}}_{0s}\!+\!p^2_s  P_{\rm FA} {\overline{P}}_{1s} \overline{P}_{\rm MD}\!+\!p^2_s P_{\rm MD}  \overline{P}_{\rm FA}{\overline{P}}_{1s}) \!\le\! 0$. Hence, $\mathcal{A}\mathcal{D}^\dagger$ and $\mathcal{D}\mathcal{A}^\dagger$ are negative semidefinite with eigenvalues $(0,0,\Lambda)$. Accordingly, $y^\dagger\mathcal{A}\mathcal{D}^\dagger y\le 0$, $y^\dagger\mathcal{D}\mathcal{A}^\dagger y\le 0$ and their sum is also negative. Based on these observations for a fixed $p_s$, the numerator of (\ref{musxxfg2}) is nonnegative\footnote{The non-negativity of the numerator and the denominator of $\mu_s$ follow from the definition of the service rate.} and concave over $\mathcal{P}$ and the denominator is positive and affine over $\mathcal{P}$; hence, $\mu_s$ is quasiconcave, as is derived in Appendix~B. Since the objective function of the optimization problem is quasiconcave and the constraints are convex for a fixed $p_s$, the problem is a quasiconcave program for each $p_s$. We solve a family of quasiconcave programs parameterized by $p_s$. The optimal $p_s$ is chosen as the one which yields the highest objective function in (\ref{opt1ofS}).

The problem of maximizing a quasiconcave function over a
convex set under convex constraints
can be efficiently and reliably solved by using the bisection method \cite{boyed}.

 Based on the construction of the dominant system $\mathcal{S}_1$ of system $\mathcal{S}$, it can be noted that the queues of the dominant system are never less than those of system $\mathcal{S}$, provided that they are both initialized identically. This is because the SU transmits dummy packets even if it does not have
any packets of its own, and therefore it always interferes with PU even if it is empty. The mean service rate of primary queue is thus reduced in the dominant system and $Q_p$ is emptied less frequently, thereby reducing also the mean service rate of the secondary queue. Given this, if the queues are stable in the dominant system, then they are stable in system $\mathcal{S}$. That is, the stability conditions of the dominant system are {\bf sufficient} for the stability of system $\mathcal{S}$. Now if $Q_s$ saturates in the dominant system, the SU will not transmit dummy packets as it always has its own packets to send. For $\boldsymbol{\lambda_p < \mu_p}$, this makes the behavior of the dominant system identical to that of system $\mathcal{S}$ and both systems are {\bf indistinguishable} at the boundary points. The stability conditions of the dominant system are thus both sufficient and {\bf necessary} for the stability of system $\mathcal{S}$ given that $\lambda_p < \mu_p$.

To get some insights for this system under the first dominant system, we consider the problem when $\lambda_p/\overline{P_p}$ is close to unity and with significant MPR capabilities,\footnote{As proposed in \cite{ourletter}, the primary parameters $\lambda_p$, $\overline{P_p}$ and $\lambda_p/\overline{P_p}$ can be efficiently estimated by overhearing the primary feedback channel.} which means that the primary queue is nonempty most of the time and therefore the optimal sensing decision is $p^*_s=0$. Note that, in general, this case provides a lower bound performance on what can be obtained in $\mathcal{S}_1$. When $p_s=0$, the maximum secondary stable throughput is given by solving the following optimization problem:
\begin{equation}
\begin{split}
& \underset{0\le p_t\le 1}{\max.} \,\,\lambda_e p_t \bigg[ \bigg(1\!-\!\frac{\lambda_p}{\mu_p}\bigg) \overline{P}_{0s}\!+\! \overline{P^{\left(\rm c\right)}_{0s}} \frac{\lambda_p}{\mu_p}\bigg],\,{\rm s.t.}\  \lambda_p\!\le\!  \mu_p
\label{opt1}
\end{split}
\end{equation}
\noindent The problem is {\bf convex} and can be solved using the Lagrangian formulation.
%Let $a\!=\!\overline{P}_{p}$, $b=\lambda_e\Delta_p$, $c=\lambda_p\big( \overline{P}_{0s}-\overline{P_{0s}^{\left({\rm c}\right)}}\big)$, and $d\!=\!\overline{P}_{0s}$.
The access probability $p_t$ is upperbounded by $\mathcal{F}$
\begin{equation}
\begin{split}
\mathcal{F}=\min\bigg\{1,\frac{\overline{P}_{p}- \lambda_p}{\lambda_e\Delta_p}\bigg\}
\end{split}
\end{equation}
The second term in $\mathcal{F}$ must be nonnegative for the problem to be feasible. The optimal access probability is thus given by
\begin{equation}
\begin{split}
p_t^*& =\min\biggr\{\mathcal{F},\max\bigg\{\frac{\overline{P}_{p}-\sqrt{\overline{P}_{p}\lambda_p\big( 1\!-\!\overline{P_{0s}^{\left({\rm c}\right)}}/\overline{P}_{0s}\big)}}{\lambda_e\Delta_p},0\bigg\}\biggr\}
\label{cvvvs}
\end{split}
\end{equation}
with $0\le\lambda_p\le \mu_p$. From the optimal solution, we notice the following remarks. As $\lambda_p$ increases, the secondary access probability, $p_t$, decreases as well. This is because the possibility of collisions increases with increasing the access probability (or increasing the secondary access to the channel) and since the PU is busy most of the time, the possibility of collisions and packet loss increase as well. In addition, by observing the optimal solution in (\ref{cvvvs}), we notice that as the secondary energy arrival, $\lambda_e$, increases, the access probability decreases. This is because accessing the channel most of the time with the availability of energy may cause high average packet loss for the PU. We note that as the capability of MPR of the primary receiver, i.e., $\overline{P_{p}^{\left({\rm c}\right)}}$, increases, the access probability of the secondary queue increases. This occurs because the possibility of decoding the primary packet under interference is almost equal to the probability of decoding without interference when the MPR capability of the primary receiver is high. Therefore, the secondary throughput increases. In addition, as the ability of the secondary receiver of decoding the secondary packets under interference, which is represented by $\overline{P_{0s}^{\left({\rm c}\right)}}/\overline{P}_{0s}$, increases, the access probability of the secondary terminal increases as far as the primary queue stability condition is satisfied.

\subsection{Second Dominant System}
In the second dominant system, denoted by $\mathcal{S}_2$, queue $Q_p$ transmits dummy packets when it is empty and the SU behaves as it would in system $\mathcal{S}$. By substituting with ${\rm Pr}\{Q_p =0\}=0$ into (\ref{dodo}), the average secondary service rate is given by

\begin{equation}
\begin{split}
\mu_s&\!=\!\Big[\overline{p_s}p_t \overline{P^{\left(\rm c\right)}_{0s}}\!+\!p_s p_f P_{\rm MD} \overline{P_{1s}^{\left({\rm c}\right)}}\!+\!p_s p_b  \overline{P_{\rm MD}}\  \overline{P_{1s}^{\left({\rm c}\right)}}\Big]{\rm Pr}\{Q_e \ne 0\}
\label{secc}
\end{split}
\end{equation}
where ${\rm Pr}\{Q_e \ne 0\}\!=\!\lambda_e$. Under this dominant system, the SU optimal sensing decision is $p^*_s=0$. This is because the PU is always nonempty. Hence, $Q_s$ mean service rate in (\ref{secc}) is rewritten as
\begin{equation}
\begin{split}
\mu_s&=p_t\lambda_e \overline{P^{\left(\rm c\right)}_{0s}}
\end{split}
\end{equation}
The probability of $Q_s$ being nonempty is $\lambda_s/\mu_s$. Hence, the primary queue mean service rate is given by
\begin{equation}
\begin{split}
\mu_p&=\bigg(1\!-\!\frac{\lambda_s}{\mu_s}\lambda_e\bigg)\overline{P}_{p}+\frac{\lambda_s}{\mu_s}\lambda_e\bigg[p_t \overline{P_{p}^{\left({\rm c}\right)}}+\left(1-p_t\right)\overline{P}_{p}\bigg]
\end{split}
\end{equation}
After some simplifications, the primary mean service rate is given by
\begin{equation}
\begin{split}
\mu_p&=\overline{P}_{p}\!-\!\frac{\lambda_s}{ \overline{P^{\left(\rm c\right)}_{0s}}}\Delta_p
\end{split}
\end{equation}
Note that $\mu_p$ is independent of $p_t$. The portion of the stability region of $\mathcal{S}$ based on $\mathcal{S}_2$ is obtained by solving a constrained optimization problem in which $\mu_p$ is maximized under the stability of the primary and the secondary queues. Since the primary mean service rate is independent of $p_t$, the stability region of the second dominant system is given by solving the following optimization feasibility problem
\begin{equation}
\begin{split}
& \underset{0\le p_t\le 1}{\max.} \,\,\overline{P}_{p}\!-\!\frac{\lambda_s}{ \overline{P^{\left(\rm c\right)}_{0s}}}\Delta_p,\, \\ & {\rm s.t.}\  \lambda_s\le p_t\lambda_e \overline{P^{\left(\rm c\right)}_{0s}}
\label{opt1}
\end{split}
\end{equation}
Hence, the optimal access probability is
\begin{equation}
\begin{split}
p_t\ge  \frac{\lambda_s}{\lambda_e \overline{P^{\left(\rm c\right)}_{0s}}}
\label{foc}
\end{split}
\end{equation}
with $\lambda_s\le \lambda_e \overline{P^{\left(\rm c\right)}_{0s}}$. Based on (\ref{foc}), the solution of the problem is a set of values which satisfies the secondary queue stability constraint. We note that as the secondary mean arrival rate, $\lambda_s$, increases, the lower limit of $p_t$ increases as well. This is because the SU must increase its service rate, which increases with the increasing of the access probability, to maintain its queue stability. We also note that one of the feasible points is $\lambda_s=\mu_s$, which means a saturated SU (since the arrival rate is equal to the service rate, ${\rm Pr}\{Q_s\!\ne\!0\}\!=\!\lambda_s/\mu_s\!=\!1$). This system is equivalent to a system with random access without employing any channel sensing with backlogged (saturated) primary and secondary transmitters.

Since the stability region of system $\mathcal{S}$ is the union of both dominant systems, the stability region of the proposed protocol always contains that of the random access without employing any spectrum sensing. Based on this observation, we can say that at high primary arrival rate, or at high probability of nonempty primary queue, the random access without employing any sensing scheme is optimal, i.e., the SU should not employ channel sensing in such case. This is because the PU is always active and therefore there is no need to sense the channel and waste $\tau$ seconds of the data transmission time.
\section{Feedback-based Access}\label{sec3}
In this section, we analyze the use of the primary feedback messages by the cognitive terminal. This system is denoted by $\Phi_{\rm F}$. In the feedback-based access scheme, the SU utilizes the available primary feedback information for accessing the channel in addition to spectrum sensing. Leveraging the primary feedback is valid when it is available and unencrypted.

In the proposed scheme, the SU monitors the PU feedback
channel. It may overhear an acknowledgment (ACK) if the primary receiver correctly decodes the primary transmission, a negative acknowledgment (NACK) if decoding fails, or nothing if there is no primary transmission. We introduce the following modification to the protocol introduced earlier in the paper. If a NACK is overheard by the SU, it assumes that the PU will retransmit the lost packet during the next time slot \cite{KarimSultan}. Being sure that the PU will be active, the secondary terminal does not need to sense the channel to ascertain the state of primary activity. Therefore, it just accesses the channel with some probability $p_r$. If an ACK is observed on the feedback channel or no primary feedback is overheard, the SU proceeds to operate as explained earlier in Section \ref{sec2}. We assume the feedback packets are very short compared to $T$ and are always received correctly by both the primary and secondary terminals due to the use of strong channel codes.

It is important to emphasize here the benefit of employing primary feedback. By avoiding spectrum sensing, the secondary terminal does not have to waste $\tau$ seconds for channel sensing. It can use the whole slot duration for data transmission. As proven in \cite{wimob,ElSh1312:Optimal}, this reduces the outage probability of the secondary link. Therefore, by differentiating between the primary states of transmission, i.e., whether they are following the reception of an ACK or not, the SU can potentially enhance its throughput by eliminating the need for spectrum sensing when the PU is about to retransmit a previously lost packet. Note that we denote the system operating exactly as system $\mathcal{S}$ with primary feedback leveraging as $\mathcal{S}_f$.
%\vspace{-0.1cm}
\subsection{First dominant system}
%\vspace{-0.2cm}
As in the previous section, under the first dominant system, denoted by $\mathcal{S}^f_1$, $Q_s$ transmits dummy packets when it is empty and the PU behaves as it would in system $\mathcal{S}_f$.
\begin{figure}
  \includegraphics[width=1\columnwidth]{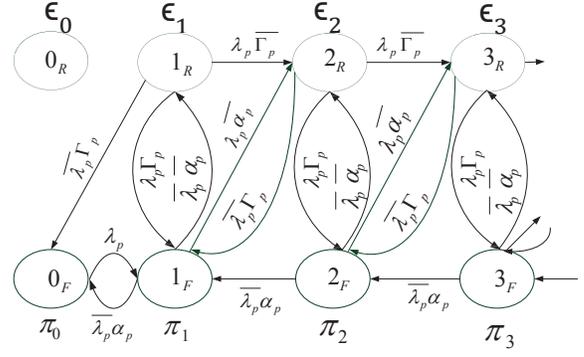}\\
  \caption{Markov chain of the PU for the feedback-based access scheme under dominant system $\mathcal{S}^f_1$. Probabilities $\overline{\Gamma_p}=1-\Gamma_p$ and $\overline{\alpha_p}=1-\alpha_p$. State self-transitions are not depicted for visual clarity.}\label{fig2}
\end{figure}

The PU's queue evolution
Markov chain under the first dominant system of this protocol is shown in Fig. \ref{fig2}. The probability of the queue
having $k$ packets and transmitting for the first time is $\pi_k$, where
$F$ in Fig. \ref{fig2} denotes first transmission. The probability of the
queue having $k$ packets and retransmitting is $\epsilon_k$, where $R$ in
Fig. \ref{fig2} denotes retransmission. Define $\alpha_p$ as the probability of successful transmission of the PU's packet in case of first transmission and $\Gamma_p$ is the probability of successful transmission of the PU's packet in case of retransmission. It can be shown that both probabilities are given by:
\begin{equation}
\begin{split}
\alpha_p&=\overline{P}_{p}\!-\!\lambda_e\Delta_p\Bigg(\overline{p_s}p_t\!+\!p_s P_{\rm MD}p_f \!+\!p_s \overline{P_{\rm MD}}p_b\Bigg)
\label{alpha}
\end{split}
\end{equation}
\begin{equation}
\begin{split}
\Gamma_p&=\overline{P}_{p}\!-\!\lambda_e\Delta_pp_r
\end{split}
\end{equation}
Solving the state balance equations, we can obtain the state probabilities which are provided in Table \ref{table}. The probability $\pi_0$ is obtained using the normalization condition $\sum_{k=0}^{\infty} (\pi_k+\epsilon_k)=1$.

It should be noticed that $\lambda_p < \eta$, where $\eta$ is defined in Table \ref{table}, is a condition for the sum $\sum_{k=0}^{\infty} (\pi_k+\epsilon_k)$ to exist. This condition ensures the existence of a stationary distribution for the Markov chain and guarantees the stability of the primary queue. The service rate of the SU is given by:
\begin{table}
\renewcommand{\arraystretch}{2}
\begin{center}
\begin{tabular}{ |c |c|  }
    \hline\hline
    $\eta$ & $\lambda_p\alpha_p+\left(1-\lambda_p\right)\Gamma_p$ \\[5pt]\hline
  $\pi_{\circ}$ & $\frac{\eta-\lambda_p}{\Gamma_p}$ \\[5pt]\hline
  $\epsilon_{\circ}$ & 0 \\[5pt]\hline
  $\pi_1$& $\pi_{\circ}\tfrac{\lambda_p}{1-\lambda_p}\frac{\lambda_p+\left(1-\lambda_p\right)\Gamma_p}{\eta}$  \\[5pt]\hline
  $\epsilon_1$ & $\pi_{\circ}\frac{\lambda_p}{\eta}\left(1-\alpha_p\right)$ \\ [5pt]\hline
  $\pi_k,k\geq 2$ & $\pi_{\circ}\frac{\lambda_p\left(1-\alpha_p\right)}{\left(1-\eta\right)^2} \bigg[ \frac{\lambda_p \left(1-\eta\right)}{\left(1-\lambda_p\right) \eta}  \bigg]^k$ \\[5pt]\hline
  $\epsilon_k,k\geq 2$ & $\pi_{\circ}\frac{\left(1-\lambda_p\right)\left(1-\alpha_p\right)}{\left(1-\eta\right)^2} \bigg[ \frac{\lambda_p \left(1-\eta\right)}{\left(1-\lambda_p\right) \eta}  \bigg]^k$ \\[5pt]\hline
  $\sum_{k=1}^{\infty}\pi_k$ & $\pi_{\circ}\frac{\lambda_p\Gamma_p}{\eta-\lambda_p}=\lambda_p$\\[5pt]\hline
   $\sum_{k=1}^{\infty}\epsilon_k$ & $\pi_{\circ}\frac{\lambda_p}{\eta-\lambda_p}\left(1-\alpha_p\right)=\frac{\lambda_p }{\Gamma_p}\left(1-\alpha_p\right)$\\[5pt]\hline
\end{tabular}
\caption{State probabilities for the feedback-based access scheme.}
\label{table}
\end{center}
\end{table}
\begin{equation}
\begin{split}
\mu_s&=\lambda_e \Bigg[\pi_0 \Bigg(\overline{p_s}p_t {\overline{P}}_{0s}+p_s p_b P_{\rm FA} {\overline{P}}_{1s}+p_s p_f  \overline{P_{\rm FA}}\ {\overline{P}}_{1s} \Bigg)\\&\,\,\,\,\,\,\,\,\,\,\,\,\,\,\,\,\,\,\ +
\Bigg(\sum_{k=1}^{\infty}\pi_k\Bigg)\Bigg(\overline{p_s}p_t \overline{P^{\left(\rm c\right)}_{0s}} \!+\!p_s p_f  P_{\rm MD} \overline{P_{1s}^{\left({\rm c}\right)}}
\\&\,\,\,\,\,\,\,\,\,\,\,\,\,\,\,\,\,\,\,\,\,\,\,\,\,\,\,\,\,\,\,\,\,\,\,\,\,\ +p_s p_b   \overline{P_{\rm MD}}\ \overline{P_{1s}^{\left({\rm c}\right)}} \Bigg)+\Bigg(\sum_{k=1}^{\infty}\epsilon_k\Bigg)p_r \overline{P_{0s}^{\left({\rm c}\right)}}\Bigg]
\end{split}
\label{muss}
\end{equation}

Let $\mathcal{I}\!=\![0,0,0,1]^\dagger$, $\mathcal{H}=-\lambda_e\Delta_p \mathcal{I}$, $\mathcal{J}=[\mathcal{D}^\dagger,0]^\dagger$, $\mathcal{\hat{U}}\!=\!-\lambda_p\mathcal{J}\!+\!\overline{\lambda_p} \mathcal{H}$, $\eta=\overline{P}_{p}+\hat{\mathcal{P}}^\dagger\mathcal{\hat{U}}$, $\hat{\mathcal{P}}=[p_t,p_f,p_b,p_r]^\dagger$, $\Gamma_p=\overline{P}_{p}+\mathcal{H}^\dagger\hat{\mathcal{P}}$, $\alpha_p=\overline{P}_{p}\!-\!\hat{\mathcal{P}}^\dagger\mathcal{J}$, and $p_r=\mathcal{I}^\dagger\hat{\mathcal{P}}$. After some algebra, and substituting by the state probabilities in Table \ref{table}, the secondary data queue mean service rate in (\ref{muss}) can be rewritten as
\begin{equation}
\begin{split}
&\mu_s \!=\! \frac{[\!(\!\overline{P_p}\!-\!\lambda_p)\mathcal{K}^\dagger\!+\!\tilde o P_p\mathcal{I}^\dagger\!+\! \lambda_p\overline{P_p}\mathcal{C}^\dagger\!] \hat{\mathcal{P}}\!+\!\hat{\mathcal{P}}^\dagger\Psi\hat{\mathcal{P}}\!}{{\overline{P_p}\!+\!\mathcal{H}^\dagger\hat{\mathcal{P}}}}
\label{bto}
\end{split}
\end{equation}
where $\mathcal{K}\!=\!\Big[\overline{p}_s {\overline{P}}_{0s},p_s  P_{\rm FA} {\overline{P}}_{1s},p_s  \overline{P_{\rm FA}}\ {\overline{P}}_{1s},0 \Big]^\dagger\!$, $\Psi\!=\!( \mathcal{\hat{U}}\mathcal{K}^\dagger \!+\!\lambda_p\mathcal{C}\mathcal{H}^\dagger\!-\!\tilde o\mathcal{J}\mathcal{I}^\dagger)$, $\mathcal{C}=\big[\overline{p_s}\ \overline{P^{\left(\rm c\right)}_{0s}},p_s P_{\rm MD} \overline{P_{1s}^{\left({\rm c}\right)}},p_s   \overline{P_{\rm MD}} \ \overline{P_{1s}^{\left({\rm c}\right)}},0 \big]^\dagger$, and $\tilde o=\lambda_p  \overline{P_{0s}^{\left({\rm c}\right)}}$. It is straightforward to show that the Hessian matrix of the numerator of (\ref{bto}) is $\nabla^2_{\hat{\mathcal{P}}} \hat{\mathcal{P}}^\dagger\Psi\hat{\mathcal{P}}\!=\! \Psi\!+\!\Psi^\dagger$ which is a negative semidefinite matrix and therefore the numerator is concave.\footnote{$\Psi$ is a negative semidefinite because it composes of three matrices $\mathcal{\hat{U}}\mathcal{K}^\dagger$, $\lambda_p\mathcal{C}\mathcal{H}^\dagger$ and $-\tilde o\mathcal{J}\mathcal{I}^\dagger$ each of which is a negative semidefinite matrix. These matrices are negative semidefinite because each of them is a nonpositive matrix (all elements are nonpositive) with rank $1$.} The denominator is affine over $\hat{\mathcal{P}}$. Since for a given $p_s$ the denominator is affine and the numerator is concave over $\hat{\mathcal{P}}$, (\ref{bto}) is quasiconcave over $\hat{\mathcal{P}}$ for each $p_s$.
%\noindent where the probability summations are given in Table \ref{table}.

 For a fixed $\lambda_p$, the maximum mean
service rate for the SU is given by solving the following
optimization problem using expression (\ref{muss}) for $\mu_s$
\begin{equation}
\begin{split}
 \underset{p_s,p_f,p_t,p_b,p_r}{\max.} \,\,\,\,\,\,\,\ & \mu_s\\
{\rm s.t.} \,\,\,\,\,\,\,\ &  0 \le p_s,p_f,p_t,p_b,p_r \le 1\\
\,\,\,\,\,\,\,\ & \lambda_p \le \eta
\end{split}
\label{190990c}
\end{equation}
The optimization problem is a quasiconcave optimization problem given $p_s$ which can be solved efficiently using the bisection method \cite{boyed}. For proof of quasiconcavity of the objective function, the reader is referred to Appendix~B. The constraint $ \lambda_p \le \eta$ is affine over $\hat{\mathcal{P}}$ for a fixed $p_s$. Since the objective function is quasiconcave given $p_s$ and the constraint is convex given $p_s$, (\ref{190990c}) is quasiconcave program for a fixed $p_s$.

Based on the construction of the dominant system $\mathcal{S}^f_1$, the queues of the dominant system are never less than those of system $\mathcal{S}_f$, provided that they are both initialized identically. This is because the SU transmits dummy packets even if it does not have
any packets of its own, and therefore it always interferes with PU even if it is empty. The mean service rate of primary queue is thus reduced in the dominant system and $Q_p$ is emptied less frequently, thereby reducing also the mean service rate of the secondary queue. Given this, if the queues are stable in the dominant system, then they are stable in system $\mathcal{S}_f$. That is, the stability conditions of the dominant system are {\bf sufficient} for the stability of system $\mathcal{S}_f$. Now if $Q_s$ saturates in the dominant system, the SU will not transmit dummy packets as it always has its own packets to send. For $\boldsymbol{\lambda_p < \eta}$, this makes the behavior of the dominant system identical to that of system $\mathcal{S}_f$ and both systems are {\bf indistinguishable} at the boundary points. The stability conditions of the dominant system are thus both sufficient and {\bf necessary} for the stability of system $\mathcal{S}_f$ given that $\lambda_p < \eta$.
\subsection{Second dominant system}
The second dominant system of $\mathcal{S}^f$ is denoted by $\mathcal{S}^f_2$. Under $\mathcal{S}^f_2$, the PU sends dummy packets when it is empty. This system reduces to a random access scheme without employing any spectrum sensing and without leveraging the primary feedback. This is because the PU is always nonempty and the optimal sensing decision is not to sense the channel at all. Moreover, the access probability of the SU is fixed over all primary states. Hence, $p^*_s=0$ and $p^*_r=p^*_t$. Accordingly, the second dominant system of $\mathcal{S}^f$ is exactly the second dominant system of $\mathcal{S}$. The stability region of system $\mathcal{S}_f$ is the union of both dominant systems.
%%%%%%%%%%%%%%%%%
%%%%%%%%%%%%%%

Following are some important notes. First, the stability region of the first dominant system of $\mathcal{S}$ or $\mathcal{S}^f$ always contains that of the second dominant system. This can be easily shown by comparing the mean service rate of nodes in each dominant system. Second, the stability region of systems $\mathcal{S}$ and $\mathcal{S}^f$ are inner bounds for the original systems $\Phi_{\rm NF}$ and $\Phi_{\rm F}$, respectively, where the energy queue is operating normally without the assumption of one packet consumption per time slot. Third, when the SU is plugged to a reliable power source, the average arrival rate is $\lambda_e=1$ packets per time slot. Under this case, the stability region of systems $\mathcal{S}$ and $\mathcal{S}^f$ coincide with their corresponding original systems $\Phi_{\rm NF}$ and $\Phi_{\rm F}$, respectively. This is because the energy queue in this case is always backlogged and never being empty regardless of the value of $\mu_e$. Hence, in general, the case of $\lambda_e=1$ energy packets per time slot is an {\bf outer bound} for the proposed systems, $\Phi_{\rm NF}$ and $\Phi_{\rm F}$, as the SU can always send data whenever its data queue is nonempty.

Next, we analyze the case of spectrum access without employing any sensing scheme to give some insights for system $\mathcal{S}^f$. Note that the results obtained for this case are tight when $\lambda_p/\overline{P_{p}}$ is close to unity and the MPR capabilities are strong. This is because, under this condition, the probability of the primary queue being empty at a given time slot is almost zero and therefore the optimal sensing decision which avoids wasting $\tau$ seconds of the transmission time is $p^*_s=0$.
\subsection{The case of $p_s=0$} \label{sec5}
Under this case, the mean service rate of the SU is given by
\begin{equation}
\begin{split}
\mu_s\!\!=\!\!\lambda_e \Bigg[\pi_0 p_t {\overline{P}}_{0s}\!+\!\Bigg(\sum_{k=1}^{\infty}\pi_k\Bigg)p_t \overline{P^{\left(\rm c\right)}_{0s}} \!+\!\Bigg(\sum_{k=1}^{\infty}\epsilon_k\Bigg)p_r \overline{P_{0s}^{\left({\rm c}\right)}}\!\Bigg]
\end{split}
\label{mussxxx}
\end{equation}
Substituting with probability of summations in Table \ref{table}, the secondary mean service rate is given by
\begin{equation}
\begin{split}
\mu_s\!\!=\!\!\lambda_e \Bigg[\pi_0 p_t {\overline{P}}_{0s}\!+\!\lambda_p p_t \overline{P^{\left(\rm c\right)}_{0s}} \!+\!\frac{\Gamma_p-\alpha_p}{\Gamma_p}p_r \overline{P_{0s}^{\left({\rm c}\right)}}\!\Bigg]
\end{split}
\label{mussxxxw}
\end{equation}

 For a fixed $\lambda_p$, the maximum
service rate for the SU is given by solving the following
optimization problem:
\begin{equation}
\begin{split}
& \underset{p_t,p_r}{\max.} \,\, \mu_s \,\ {\rm s.t.} \,\,\,\,\  0 \le p_t,p_r \le 1,\,\ \lambda_p \le \eta
\end{split}
\end{equation}
\noindent The optimization problem is quasiconcave (quasiconcave objective with a linear constraint) and can be solved using bisection method \cite{boyed}. Fixing $p_r$ makes the optimization problem a convex program parameterized by $p_r$. The optimal $p_r$ is taken as that which yields the highest value of the objective function.
Let $\ell\!=\!\lambda_e\Delta_p$.
We obtain the following optimization problem for a given $p_r$:
\begin{equation}
\begin{split}
 \underset{\substack{{0\le p_t\le 1}}}{\max.} \,\,\,\,\,\ &\bigg(\frac{\overline{P_p}\!-\!\overline{\lambda_p}\ell p_r\!-\!\lambda_p\!+\!\lambda_p \frac{\overline{P^{\left(\rm c\right)}_{0s}}}{{{\overline{P}}_{0s}}}\overline{P_p}}{\Gamma_p}\bigg)p_t\!-\!\frac{\lambda_p \ell }{\Gamma_p} p_t^2  \\
{\rm s.t.}  \,\,\,\,\,\ & \lambda_p \le \eta
% \Longleftrightarrow p_t\le \frac{\overline{P_p}-\overline{\lambda_p}\ell p_r-\lambda_p}{\lambda_p\ell}
\end{split}
\label{opt15}
\end{equation}
The objective function of (\ref{opt15}) is concave over convex set under linear constraints and therefore a concave program. It can be solved using the Lagrangian formulation. Setting the first derivative of the objective function to zero, the root of the first derivative is given by
\begin{equation}
\begin{split}
p_t\!=\!\frac{{\overline{P_p}\!-\!\overline{\lambda_p}\ell p_r\!-\!\lambda_p\!+\!\lambda_p \frac{\overline{P^{\left(\rm c\right)}_{0s}}}{{{\overline{P}}_{0s}}}\overline{P_p}}}{2{\lambda_p \ell }}
\end{split}
\end{equation}
Since $\lambda_p \le \eta=\lambda_p \alpha_p\!+\!\overline{\lambda_p} \Gamma_p$ and using (\ref{alpha}), the access probability is upperbounded as
%\begin{equation}
%\begin{split}
%\lambda_e [p_t (\overline{P_{p}^{\left({\rm c}\right)}}-\overline{P}_{p})]+\overline{P}_{p}=\alpha_p
%\end{split}
%\end{equation}
\begin{equation}
\begin{split}
p_t\le \frac{\overline{P_p}-\overline{\lambda_p}\ell p_r-\lambda_p}{\lambda_p\ell}
\end{split}
\end{equation}
The optimal solution is then given by
\begin{equation}
\begin{split}
p_t^*\!=\!\min\Bigg\{\frac{\overline{P_p}\!-\!\overline{\lambda_p}\ell p_r\!-\!\lambda_p}{\lambda_p\ell}, \frac{{\overline{P_p}\!-\!\overline{\lambda_p}\ell p_r\!-\!\lambda_p\!+\!\lambda_p \frac{\overline{P^{\left(\rm c\right)}_{0s}}}{{{\overline{P}}_{0s}}}\overline{P_p}}}{2{\lambda_p \ell }}\Bigg\}
\label{pt1}
\end{split}
\end{equation}
%Note that the optimization problems (\ref{opt1}) and (\ref{opt15}) are solved at the SU.
\section{Delay-Aware Primary Users}
\label{delayaware}
In this section, we investigate the primary queueing delay and plug a constraint on the primary queueing delay to the optimization problems. That is, we maximize $\mu_s$ under the constraints that the primary queue is stable and that the primary packet delay is smaller than or equal a specified value $D\!\ge\!1$.\footnote{Note that based on the adopted arrival model, the minimum primary queueing delay is $1$ time slot, i.e., $D=1$ time slot.} The value of $D\!\ge\!1$ is application-dependent and is related to the required QoS for the PU. Delay analysis for interacting queues is a notoriously hard problem \cite{sadek}. To bypass this difficulty, we consider the special case where the secondary data
queue is always backlogged (or saturated)\footnote{This case equivalent to the first dominant systems of $\mathcal{S}$ and $\mathcal{S}^f$.} while the primary queue behaves exactly as it would in the original systems $\Phi_{\rm NF}$ and $\Phi_{\rm F}$. This
represents a lower bound (or worst-case scenario) on performance for the PU compared
with the original systems in which the secondary data queue is not backlogged all the time. Next, we compute the primary queueing delay under each system.
\subsection{Primary Queueing Delay for System $\mathcal{S}$ with Saturated SU}
Let $D_p$ be the average delay of the primary queue. Using Little's law and (\ref{nuk}),
\begin{equation}
\begin{split}
D_p=\frac{1}{\lambda_p}\sum_{k=1}^{\infty}k\nu_k=\frac{1-\lambda_p}{\mu_p-\lambda_p}
\end{split}
\label{190990}
\end{equation}
\noindent For the optimal random access and sensing, we solve the following constrained optimization problem. We maximize the mean secondary service rate under the constraints that the primary queue is stable and that the primary packet delay is smaller than or equal a specified value $D$. The optimization problem with
 $\mu_p$ given in (\ref{mup}) and $\mu_s$ in (\ref{mus}) can be written as
\begin{equation}
\begin{split}
& \max_{p_s,p_f,p_b,p_t} \,\,\mu_s\\
&\,\,{\rm s.t.} \,\,\,\,\  0 \le p_s,p_f,p_b,p_t \le 1\\
&\,\,\,\,\,\,\,\,\,\,\,\,\,\,\, \lambda_p\le \mu_p\\& \,\,\,\,\,\,\,\,\,\,\,\,\,\,D_p \le D
\end{split}
\label{190990}
\end{equation}
The delay constraint in case of system $\mathcal{S}$ with backlogged SU can be converted to a constraint on the primary mean service rate. That is, $D_p=\frac{1-\lambda_p}{\mu_p-\lambda_p}\!\le\!  D$ can be rewritten as $\mu_p \ge \lambda_p+\frac{1-\lambda_p}{D}$. The intersection of the stability constraint and the delay constraint is the delay constraint. That is, the set of $\lambda_p$ which satisfies the delay constraint is $\{\lambda_p:   \lambda_p\le\mu_p-\frac{1-\lambda_p}{D}\}=\{\lambda_p\le\frac{\mu_p-1/D}{(1-1/D)}\le \mu_p\}$, whereas the set of $\lambda_p$ which satisfies the stability constraint is  $\{\lambda_p: \lambda_p \le \mu_p\}$. The intersection of both sets is given by $\{\lambda_p\le\frac{\mu_p-1/D}{(1-1/D)}\} \cap \{\lambda_p: \lambda_p \le \mu_p\}= \{\lambda_p:   \lambda_p\le\frac{\mu_p-1/D}{(1-1/D)}\}$, both sets are equal when the delay constraint approaches $\infty$, i.e., $D\rightarrow \infty$. Hence, the delay constraint subsumes the stability constraint.

The optimization problem is quasiconcave given $p_s$ because $\mu_s$ is quasiconcave (the proof is in Appendix~B) and the delay constraint is linear on the optimization parameters.

At high $\lambda_p/\overline{P_{p}}$ and strong MPR capabilities, the probability of the primary queue being empty at a given time slot is almost zero and therefore the optimal sensing decision is $p^*_s=0$. In this case, we can get the optimal solution of the optimization problem.
The optimization problem can be stated as
\begin{equation}
\begin{split}
& \underset{0\le p_t\le 1}{\max.} \,\,\lambda_e p_t \bigg[ \bigg(1\!-\!\frac{\lambda_p}{\mu_p}\bigg) \overline{P}_{0s}\!+\! \overline{P^{\left(\rm c\right)}_{0s}} \frac{\lambda_p}{\mu_p}\bigg],\,{\rm s.t.}\  \lambda_p\!\le\!  \mu_p,\ D_p \!\le\! D
\label{opt1}
\end{split}
\end{equation}
where
\begin{equation}
\begin{split}
\mu_p&=\overline{P}_{p}\!-\!\lambda_e\Delta_pp_t
\end{split}
\end{equation}
\noindent The optimization problem (\ref{opt1}) is {\bf convex} and can be solved using the Lagrangian formulation. The delay constraint subsumes the stability constraint, $\lambda_p < \mu_p$, and $p_t$ is upperbounded by $\mathcal{U}$
\begin{equation}
\begin{split}
\mathcal{U}=\min\bigg\{1,\frac{\overline{P}_{p}-\bigg(\frac{1-\lambda_p}{D}+ \lambda_p\bigg)}{\lambda_e\Delta_p}\bigg\}
\end{split}
\end{equation}
The second term in $\mathcal{U}$ must be nonnegative for the problem to be feasible. The optimal access probability is thus given by
\begin{equation}
\begin{split}
p_t^*& =\min\biggr\{\mathcal{U},\max\bigg\{\frac{\overline{P}_{p}\!-\!\sqrt{\overline{P}_{p}\lambda_p\big( 1\!-\!\overline{P_{0s}^{\left({\rm c}\right)}}/\overline{P}_{0s}\big)}}{\lambda_e\Delta_p},0\bigg\}\biggr\}
\label{poor}
\end{split}
\end{equation}
 From the optimal solution (\ref{poor}), we can establish here a similar argument about the impact of each parameter on the secondary access probability as the one beneath (\ref{pt1}). However, the difference here is that we have the impact of the delay constraint which has the following affect on the secondary access probability. As the delay constraint, $D$, increases, the access probability of the SU decreases to avoid increasing collisions with the PU which causes primary throughput loss. If the amount of collisions is high, the delay constraint may be violated if the SU accesses with an access probability higher than $p_t^*$.
\subsection{Primary Queueing Delay for System $\mathcal{S}^f$ with Saturated SU}
Applying Little's law, the primary queueing delay is given by
\begin{equation}
D_p=\frac{1}{\lambda_p}\sum_{k=1}^{\infty} k \left(\pi_k+\epsilon_k\right)
\end{equation}
\noindent Using the state probabilities provided in Table \ref{table},
\begin{equation}
\begin{split}
D_p &=\frac{(\alpha_p-\eta)(\eta-\lambda_p)^2+\left(1-\lambda_p\right)^2 \left(1-\alpha_p\right) \eta}{(\eta-\lambda_p)\left(1-\lambda_p\right) \left(1-\eta\right) \Gamma_p}
  \end{split}
\end{equation}
%It can be shown that $D_p$ is a convex function over $\eta$ and hence over $\mathcal{P}$.
 For a fixed $\lambda_p$, the maximum mean
service rate for the SU is given by solving the following
optimization problem using expression (\ref{muss}) for $\mu_s$
\begin{equation}
\begin{split}
 \max_{p_s,p_f,p_t,p_b,p_r} \,\,\mu_s&\\
\,\,{\rm s.t.} \,\,\,\,\  0 &\le p_s,p_f,p_t,p_b,p_r \le 1\\
\,\,\,\ \lambda_p &\le \eta
\\
\,\,\,\ D_p &\le D
\end{split}
\label{190990}
\end{equation}
Note that $\mu_s$ is given in Eqn. (\ref{bto}). The optimization problem can be shown to be a concave program for a given $p_s$ and $p_r$. More Specifically, for a fixed $p_r$, the denominator in (\ref{bto}) becomes a constant. Since the numerator is concave for a given $p_s$ as shown beneath (\ref{bto}), the objective function of (\ref{190990}) is concave. The delay constraint can be rewritten as

\begin{equation}
\begin{split}
 \mathcal{E}\!=\! \frac{(\eta\!-\!\Gamma_p)(\eta\!-\!\lambda_p)^2}{\eta}\!+\!\!\overline{\lambda_p} (\mathcal{W}\!-\!\eta)  \!+\!  \frac{D(\eta\!-\!\lambda_p) (\eta\!-\!1)\Gamma_p \lambda_p}{\eta}  \!\le\! 0
\end{split}
\end{equation}
\noindent where $\mathcal{W}\!=\!\lambda_p\!+\!\overline{\lambda_p} \Gamma_p$ and $D \!\ge\! 1$. The second derivative of $\mathcal{E}$ for a given $p_r$ with respect to $\eta$ is given by

\begin{equation}
\begin{split}
 \nabla^2_\eta \mathcal{E}\!=\! \frac{2(\Gamma_p (D-1) \lambda_p^2+\eta^3)}{\eta^3}\ge 0
\end{split}
\end{equation}
$\nabla^2_\eta \mathcal{E}$ is always nonnegative. Hence, $\mathcal{E}$ is convex over $\eta$ for a fixed $p_r$. Since $\eta$ is affine over $\hat{\mathcal{P}}$ for a fixed $p_s$, $\mathcal{E}$ is then convex over $\hat{\mathcal{P}}$. This completes the proof of concavity of the optimization problem (\ref{190990}) for a given $p_s$ and $p_r$. Note that we solve a family of concave problems parameterized by $p_s$ and $p_r$. The optimal pair $(p_r,p_s)$ is taken as the pair which yields the highest objective function in (\ref{190990}).
\section{Numerical Results and Conclusions}\label{numerical}
In this section, we provide some numerical results for the optimization problems presented in this paper. A random access without employing spectrum sensing is simply obtained from system $\mathcal{S}$ by setting $p_s$ to zero. Let $\mathcal{S}_R$ and $\mathcal{S}^f_R$ denote the random access system without employing any spectrum sensing without and with feedback leveraging, respectively.
We also introduce the conventional scheme of spectrum access, denoted by $\mathcal{S}^c$. In this system, the SU senses the channel each time slot for $\tau$ seconds. If the PU is sensed to be inactive, the SU accesses with probability $1$. If the PU is sensed to be active, the SU remains silent. The mean service rates for this case are obtained from Section \ref{sec2} with $p_t=0$, $p_s=1$, $p_f=1$ and $p_b=0$. We define here two variables $\delta_{0s}=\frac{{\overline{P}}^{\left({\rm c}\right)}_{0s}}{{\overline{P}}_{0s}}$ and $\delta_{1s}=\frac{\overline{P_{1s}^{\left({\rm c}\right)}}}{{\overline{P}}_{1s}}$, both of them are less than $1$ as shown in Appendix A. Fig.\ \ref{fig1x} shows the stability region of the proposed protocols. Systems $\mathcal{S}_R$ and $\mathcal{S}^f_R$ are also plotted. The parameters used to generate the figure are: $\lambda_e=1$ energy packets/slot, $\overline{P}_{p}=0.7$, $\overline{P_{p}^{\left({\rm c}\right)}}=0.1$, $\overline{P}_{0s}=0.8$, $\overline{P^{\left(\rm c\right)}_{0s}}=0.1$, $\overline{P}_{1s}=0.6$, $\overline{P_{1s}^{\left({\rm c}\right)}}=0.3$, $P_{\rm FA}=0.01$, and $P_{\rm MD}=0.02$. We can note that primary feedback leveraging expands the stability region. It is also noted that randomly accessing the channel without channel sensing and with primary feedback leveraging can outperform system $\mathcal{S}$ for some $\lambda_p$. This is because in system $\mathcal{S}_R^f$ the SU does not sense the channel at the following time slot to primary packet decoding failure at the primary destination. Therefore, the SU does not waste $\tau$ seconds in channel sensing and it is sure of the activity of the PU.

 Fig. \ref{fig2x} provides a comparison between the maximum secondary stable throughput for the proposed systems and the conventional system. The parameters used to generate the figure are: $\lambda_e=0.4$ energy packets/slot, $\overline{P}_{p}=0.7$, $\overline{P_{p}^{\left({\rm c}\right)}}=0.1$, $\overline{P}_{0s}=0.8$, $\overline{P^{\left(\rm c\right)}_{0s}}=0.1$, $\overline{P}_{1s}=0.6$, $\overline{P_{1s}^{\left({\rm c}\right)}}=0.075$, $P_{\rm FA}=0.05$, and $P_{\rm MD}=0.01$. For the investigated parameters, over $\lambda_p\!<\! 0.475$ packets/slot, the proposed protocols outperform the conventional system. Whereas over $\lambda_p\!\ge\! 0.475$ packets/slot, all systems provide the same performance. This is because at high primary arrival rate, the probability of the primary queue being empty is very low and the PU will be active most of the time slots. Hence, the SU senses the channel each time slot and avoids accessing the channel when the PU is sensed to be active and at retransmission states. That is, $p_t=0$, $p_s=1$, $p_r=0$, $p_b=0$ and $p_f=1$. We note that feedback leveraging always enhances the secondary throughput.

 Figs.\ \ref{fig3x} and  \ref{fig6x} show the impact of the MPR capability at the receiving nodes on the stable throughput region. Without MPR capability, collisions are assumed to lead to sure packet loss. Therefore, a collision model without MPR corresponds to the case of the probabilities of correct reception being zero when there are simultaneous transmissions. As shown in Fig. \ref{fig3x}, the secondary service rate is reduced when there is no MPR capability. As the strength of MPR capability increases, the stability regions expand significantly. It can be noted that the performance of $\mathcal{S}$ and $\mathcal{S}^f$ are equal when the MPR capability is high. This is due to the fact that the SU does not need to employ channel sensing or feedback leveraging as it can transmit each time slot simultaneously with the PU because the secondary receiver can decode packets under interference with a probability almost equal to the probability when it transmits alone. The figure is plotted for different MPR strength of the secondary receiver, namely, for $P_1\!=\!\delta_{0s}\!=\!\delta_{1s}\!=\!0$, $P_2\!=\!\delta_{0s}\!=\!\delta_{1s}\!=\!1/8$, $P_3=\delta_{0s}\!=\!\delta_{1s}\!=\!1/4$ and $P_4\!=\!\delta_{0s}\!=\!\delta_{1s}\!=\!1/2$. The parameters used to generate the figure are: $\lambda_e=0.4$ energy packets/slot, $\overline{P}_{p}=0.7$, $\overline{P_{p}^{\left({\rm c}\right)}}=0.1$, $\overline{P}_{0s}=0.8$, $\overline{P}_{1s}=0.6$, $P_{\rm FA}=0.05$, and $P_{\rm MD}=0.01$. Fig. \ \ref{fig6x} demonstrates the impact of the MRR capability of the primary receiver on the stability region of system $\mathcal{S}$. As can be seen, the increases of $\overline{P}_p^{\left({\rm c}\right)}$ increases the secondary stable throughput for each $\lambda_p$. The parameters used to generate the figure are: $\lambda_e=0.8$ energy packets/slot, $\overline{P}_{p}=0.7$, $\overline{P}_{0s}=0.8$, $\overline{P^{\left(\rm c\right)}_{0s}}=0.1$, $\overline{P}_{1s}=0.6$, $\overline{P_{1s}^{\left({\rm c}\right)}}=0.075$, $P_{\rm FA}=0.05$, and $P_{\rm MD}=0.01$ and for different values of $\overline{P}_p^{\left({\rm c}\right)}$.

  Fig.\ \ref{fig4x} shows the impact of the energy arrival rate on the secondary stable throughput for the considered systems. The parameters used to generate the figure are: $\lambda_p=0.4$ packets/slot, $\overline{P}_{p}=0.7$, $\overline{P_{p}^{\left({\rm c}\right)}}=0.1$, $\overline{P}_{0s}=0.8$, $\overline{P^{\left(\rm c\right)}_{0s}}=0.1$, $\overline{P}_{1s}=0.6$, $\overline{P_{1s}^{\left({\rm c}\right)}}=0.075$, $P_{\rm FA}=0.05$, and $P_{\rm MD}=0.01$. As expected, the secondary service rate increasing with increasing $\lambda_e$. We note that there are some constant parts in systems $\mathcal{S}_R$ and $\mathcal{S}^f_R$ at high $\lambda_e$. This is due to the fact that increasing the energy arrivals at the energy queue may not boost the secondary throughput because the SU even if it has a lot of energy packets it cannot violate the primary QoS. The violation of the primary QoS may occur due to the presence of sensing errors. We also note that at low energy arrival rate, all systems have the same performance. This is because the secondary access probabilities and the rate in each system are limited by the mean arrival rate of the secondary energy arrival rate. Fig.\ \ref{fig5x} demonstrates the impact of varying the primary queueing delay constraint, $D$, on the secondary service rate. The parameters used to generate the figure are: $\lambda_e=0.4$ energy packets/slot, $\overline{P}_{p}=0.7$, $\overline{P_{p}^{\left({\rm c}\right)}}=0.1$, $\overline{P}_{0s}=0.8$, $\overline{P^{\left(\rm c\right)}_{0s}}=0.1$, $\overline{P}_{1s}=0.6$, $\overline{P_{1s}^{\left({\rm c}\right)}}=0.075$, $P_{\rm FA}=0.05$, and $P_{\rm MD}=0.01$ and two different values of the primary queueing delay constraint. As is clear from the figure, the secondary service rate is reduced when the primary queueing delay constraint is more strict.

\begin{figure}
  \includegraphics[width=1\columnwidth]{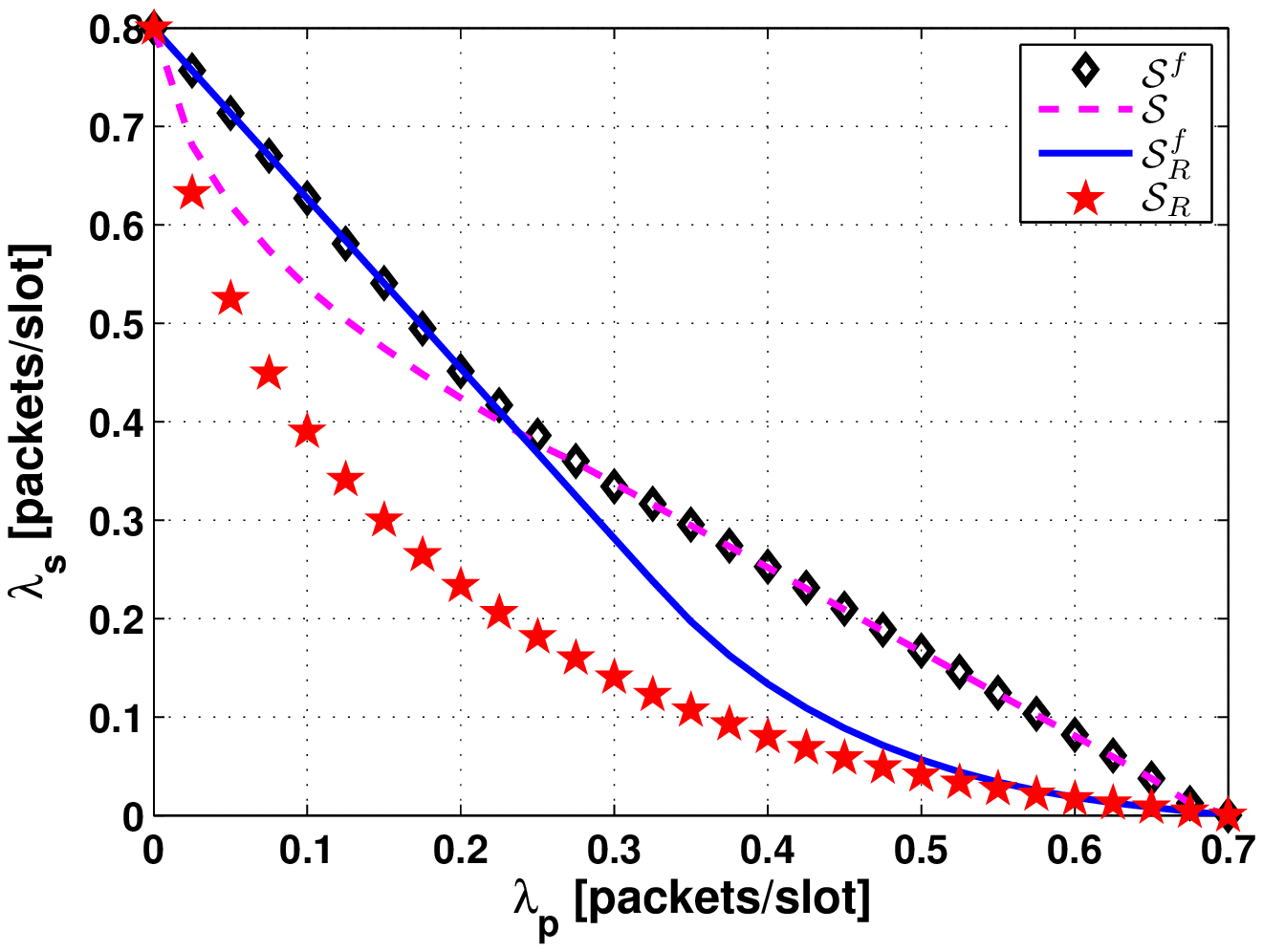}\\
   \caption{Stability region of the proposed systems.}\label{fig1x}
\end{figure}
\begin{figure}
   \includegraphics[width=1\columnwidth]{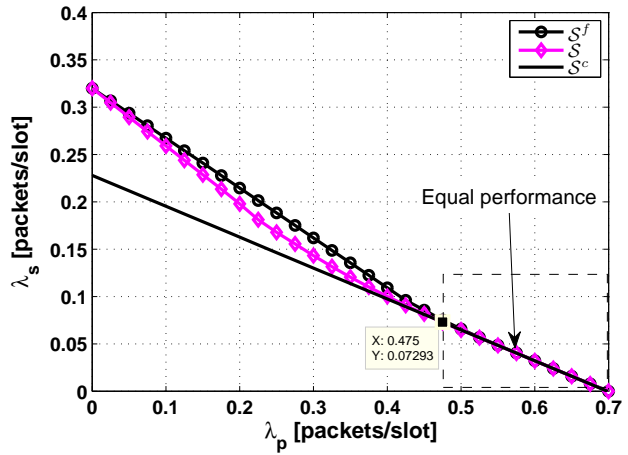}\\
   \caption{Stability region of the proposed systems. The conventional system, $\mathcal{S}^c$, is also plotted for comparison purposes.}\label{fig2x}
\end{figure}

\begin{figure}
  \includegraphics[width=1\columnwidth]{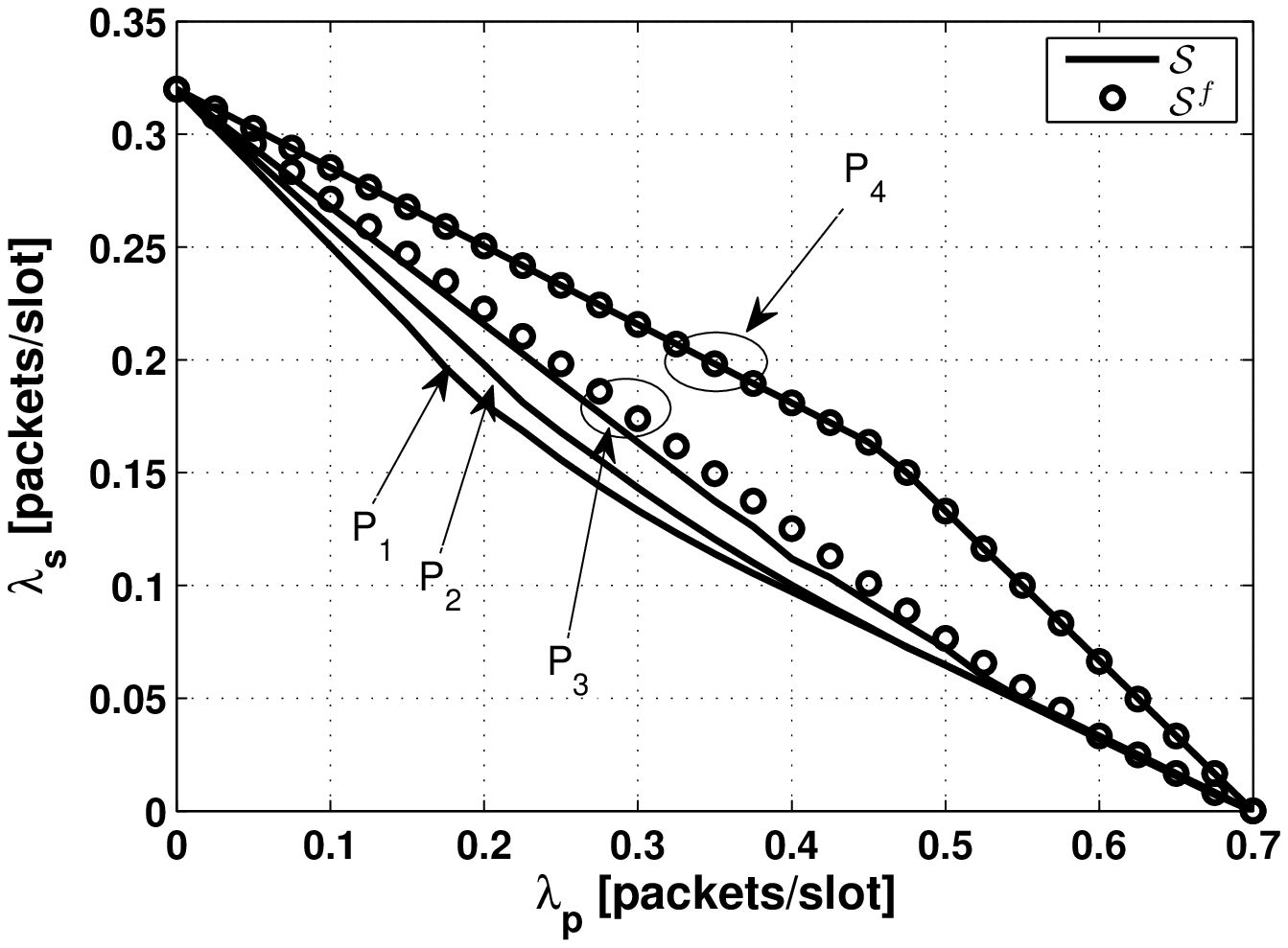}\\
   \caption{Stability region of the proposed systems.}\label{fig3x}
\end{figure}

\begin{figure}
  \includegraphics[width=1\columnwidth]{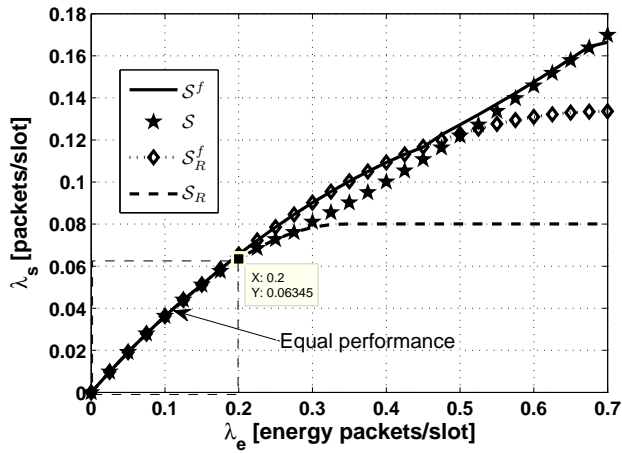}\\
   \caption{Stability region of system $\mathcal{S}$ for different values of the primary receiver MPR capability.}\label{fig6x}
\end{figure}

\begin{figure}
%  % Requires \usepackage{graphicx}
  \includegraphics[width=1\columnwidth]{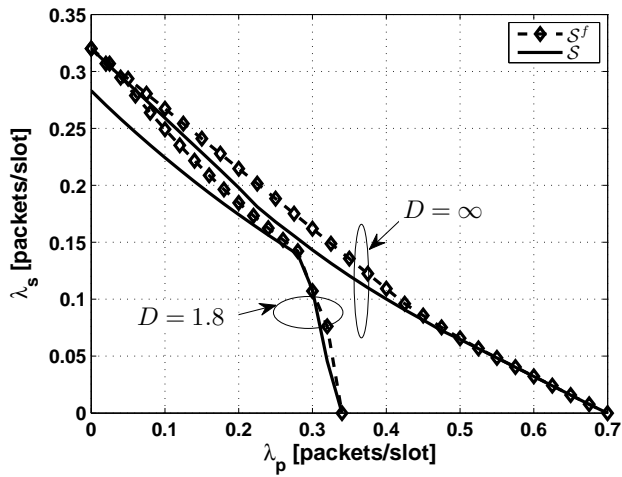}\\
   \caption{Maximum secondary throughput versus energy arrival rate.}\label{fig4x}
\end{figure}

\begin{figure}
  \includegraphics[width=1\columnwidth]{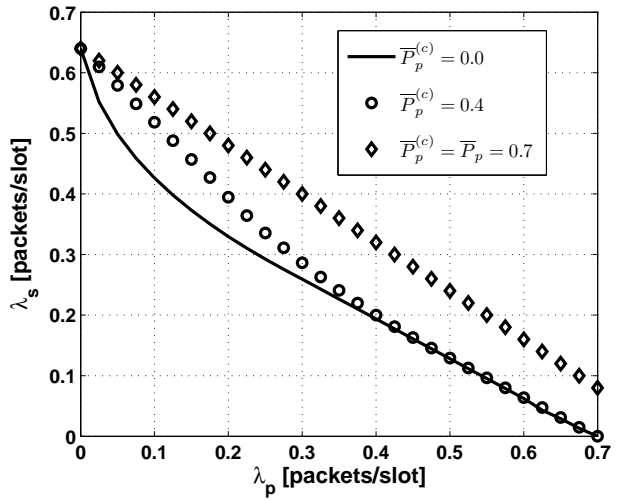}\\
   \caption{Maximum secondary throughput versus $\lambda_p$ for specific primary queueing delay.}\label{fig5x}
\end{figure}
\section*{Appendix A}
We adopt a flat fading channel model and assume that the channel gains remain constant over the duration of the time slot. We do not assume the availability of transmit channel state information (CSI) at the transmitters. Each receiver is modeled as zero mean additive white Gaussian noise (AWGN). We derive here a generic expression for the outage probability at the receiver of transmitter $j$ (node $k$) when there is concurrent transmission from the transmitter $v$. Assume that node $j$ starts transmission at $i\tau$ and node $v$ starts transmission at $n\tau$. Outage occurs when the spectral efficiency  $R^{\left(i\right)}_j\!=\!\frac{b}{WT^{\left(i\right)}_j}$, where $W$ is the channel bandwidth, $T^{\left(i\right)}_j$ is the transmission time of node $j$ and $b$ is number of bits per data packet, exceeds the channel capacity
\begin{equation}
P_{jk,in}^{\left({\rm c}\right)}={\rm Pr}\biggr\{R^{\left(i\right)}_j >  \log_{2}\left(1+\frac{\gamma_{jk,i} \beta_{jk}}{\gamma_{vk,n} \beta_{vk}+1}\right)\biggr\}
\end{equation}
\noindent where the superscript ${\rm c}$ denotes concurrent transmission, ${\rm Pr}\{.\}$ denotes the probability of the argument, $\beta_{jk}$ is the channel gain of link  $j\rightarrow k$, $\mathcal{N}_k$ is the noise variance at receiver $k$ in Watts, $\gamma_{jk,i}\!=\!\mathbb{P}^{\left(i\right)}_j/\mathcal{N}_k$, $\mathbb{P}^{\left(i\right)}_j$ Watts is the transmit power employed by node $j$ when it starts transmission at $t\!=\!i\tau$, $\gamma_{vk,n}\!=\!\mathbb{P}^{\left(n\right)}_\nu/\mathcal{N}_k$, and $\mathbb{P}^{\left(n\right)}_v$ is the used transmit power by node $v$ when it starts transmission at $t\!=\!n\tau$.
The outage probability can be written as
\begin{equation}\label {1900}
P_{jk,in}^{\left({\rm c}\right)}={\rm Pr}\Big\{\frac{\gamma_{jk,i} \beta_{jk}}{\gamma_{vk,n} \beta_{vk}+1}<{2^{R^{\left(i\right)}_j}-1}\Big\}
\end{equation}
\noindent Since $\beta_{jk}$ and $\beta_{vk}$ are independent and exponentially distributed (Rayleigh fading channel gains) with means $\sigma_{jk}$ and $\sigma_{vk}$, respectively, we can use the probability density functions of these two random variables to obtain the outage probability of link $j\rightarrow k$ as
 \begin{eqnarray}\label{193}
 P_{jk,in}^{\left({\rm c}\right)}=1-\frac{1}{1+\Big( {2^{R^{\left(i\right)}_j}-1} \Big)\frac{\gamma_{vk,n}\sigma_{vk}}{ \gamma_{jk,i}\sigma_{jk}}} {\exp\Big(-\frac{{2^{R^{\left(i\right)}_j}-1}}{\gamma_{jk,i} \sigma_{jk}}\Big)}
\end{eqnarray}
We note that from the outage probability (\ref{193}), the numerator is increasing function of $R^{\left(i\right)}_j$ and the denominator is a decreasing function of $R^{\left(i\right)}_j$. Hence, the outage probability $P_{jk,in}^{\left({\rm c}\right)}$ increases with $R^{\left(i\right)}_j$.
The probability of correct reception $\overline{P^{\left({\rm c}\right)}_{jk,i}}=1-P^{\left({\rm c}\right)}_{jk,i}$ is thus given by

  \begin{eqnarray}\label{conctra}
 \overline{P_{jk,in}^{\left({\rm c}\right)}}=\frac{\overline{P}_{jk,i}}{1+\Big({2^{\frac{b}{TW\left(1-\frac{i\tau}{T}\right)}}-1} \Big)\frac{\gamma_{vk,n}\sigma_{vk}}{\gamma_{jk,i} \sigma_{jk}}}\le\overline{P}_{jk,i}
\end{eqnarray}
\noindent where $\overline{P}_{jk,i}\!=\!\exp\Big(-\frac{{2^{R^{\left(i\right)}_j}-1}}{\gamma_{jk,i} \sigma_{jk}}\Big)$ is the probability of packet correct decoding at receiver $k$  when node $j$ transmits alone (without interference). As is obvious, the probability of correct reception is lowered in the case of interference.
%%%%

Following are some important notes. Firstly, note that if the PU's queue is nonempty, the PU transmits the packet at the head of its queue at the beginning of the time slot with a fixed transmit power $\mathbb{P}_{\rm p}$ and data transmission time $T_{\rm p}\!=\!T$. Accordingly, the superscript $i$ which represents the instant that a transmitting node starts transmission in is removed in case of PU.

Secondly, for the SU, the formula of probability of complement outage of link ${\rm s}\rightarrow {\rm sd}$ when the PU is active is given by
  \begin{eqnarray}\label{stconctra}
 \overline{P_{{\rm s} ,i0}^{\left({\rm c}\right)}}=\frac{\exp\Big(-\frac{{2^{\frac{b}{TW\left(1-\frac{i\tau}{T}\right)}}-1}}{\gamma_{{\rm s} {\rm sd},i} \sigma_{{\rm s} {\rm sd}}}\Big)}{1+\Big({2^{\frac{b}{TW\left(1-\frac{i\tau}{T}\right)}}-1} \Big)\frac{\gamma_{{\rm p} {\rm sd},0}\sigma_{{\rm p} {\rm sd}}}{\gamma_{{\rm s} {\rm sd},i} \sigma_{{\rm s} {\rm sd}}}}\!
\end{eqnarray}
where $n\!=\!0$ because the PU always transmits at the beginning of the time slot and $\gamma_{{\rm s} {\rm sd},i}~=~{\rm e}/(T(1~-~i\tau/T)\mathcal{N}_{sd})~=~\gamma_{{\rm s} {\rm sd},0}/(1~-~i\tau/T)$. The denominator of (\ref{stconctra}) is proportional to $\Big({2^{\frac{b}{TW\left(1-\frac{i\tau}{T}\right)}}-1} \Big) (1-i\frac{\tau}{T})$, which in turn is monotonically decreasing with $i\tau$. Using the first derivative with respect to $i\tau$, the numerator of (\ref{stconctra}), $\overline{P}_{{\rm s}{\rm },i0}~=~\exp\Big(\!-\!\frac{{2^{\frac{b}{TW\left(1-\frac{i\tau}{T}\right)}}-1}}{\frac{\rm e}{T(1-i\frac{\tau}{T})} \sigma_{{\rm s} {\rm sd}}}\Big)$, can be easily shown to be decreasing with $i\tau$ as in \cite{wimob,ElSh1312:Optimal}. Since the numerator of (\ref{stconctra}) is monotonically decreasing with $i\tau$ and the denominator is monotonically increasing with $i$, $ \overline{P_{{\rm s},i0}^{\left({\rm c}\right)}}$ is monotonically decreasing with $i\tau$. Therefore, the secondary access delay causes reduction in probability of secondary packets correct reception at the secondary destinations.

Thirdly, for the PU, $i=0$, $j=p$ and $k=pd$, the formula of probability of complement outage of link ${\rm p}\rightarrow {\rm pd}$ when the SU transmits at $n\tau$ is given by
  \begin{eqnarray}
 \overline{P_{{\rm p},0n}^{\left({\rm c}\right)}}=\frac{\overline{P}_{{\rm p},0}}{1+\Big({2^{\frac{b}{TW }}-1} \Big)\frac{\gamma_{{\rm s}{\rm pd},n}\sigma_{{\rm s}{\rm pd}}}{\gamma_{{\rm p}{\rm pd},0} \sigma_{{\rm p}{\rm pd}}}}
 \label{ccdfc}
\end{eqnarray}
Since $\tau/T \ll 1$, $\gamma_{{\rm s}{\rm pd},n}=\frac{e}{T(1-n\tau/T)\mathcal{N}_{pd}}$ for $n\in\{0,1\}$ is then approximately given by $\gamma_{{\rm s}{\rm pd},n}=\gamma_{{\rm s}{\rm pd}}\!=\!\frac{e}{T\mathcal{N}_{pd}}$. Hence, the impact of $\tau$ or secondary access delay on the primary outage probability is insignificant and it can be eliminated. That is,
  \begin{eqnarray}
 \overline{P_{{\rm p},0n}^{\left({\rm c}\right)}}\approx \frac{\overline{P}_{{\rm p},0}}{1+\Big({2^{\frac{b}{TW }}-1} \Big)\frac{\gamma_{{\rm s}{\rm pd}}\sigma_{{\rm s}{\rm pd}}}{\gamma_{{\rm p}{\rm pd},0} \sigma_{{\rm p}{\rm pd}}}}= \overline{P_{{\rm p}}^{\left({\rm c}\right)}}
\end{eqnarray}

Based on the above, we simply denote the probability of correct reception for the PU without and with interference as $\overline{P_{{\rm p}}}$ and $\overline{P_{{\rm p}}^{\left({\rm c}\right)}}$, respectively. The probability of correct reception for the SU without and with interference when it starts transmission from $i\tau$ seconds relative to the beginning of the time slot are denoted by $\overline{P_{i{\rm s}}}$ and $\overline{P_{i{\rm s}}^{\left({\rm c}\right)}}$, respectively.
\section*{Appendix B}
In this Appendix, we prove the quasiconcavity of $V(\rho)=\theta(\rho)/w(\rho)$, where $\theta(\rho)$ is nonnegative and concave, $w(\rho)$ is positive and affine, and $\rho=[\rho_1,\rho_2,\dots,\rho_\mathcal{M}]^\dagger$, $\mathcal{M}$ is a positive integer, belongs to the compact set ${\bf dom} V(\rho)=[0,1]^\mathcal{M}$ which is a convex set. Let $S_\zeta$ to be the $\zeta$-superlevel set of $V(\rho)$ which is given
by $S_\zeta = \{\rho \in {\bf dom} V(\rho) | V(\rho)\ge \zeta \}$. The quasiconcavity of $V(\rho)$ is proved as follows. Since $\theta(\rho)\ge 0$ and $w(\rho)>0$, it suffices to show that $S_\zeta$ are convex sets for all $\zeta\in \mathbb{R}$, $\mathbb{R}$ is the set of real numbers \cite{boyed}. If $\zeta <0$, then by the non-negativity of $V(\rho)$, we have
$S_\zeta= \{\rho \in {\bf dom} V(\rho) | V(\rho)\ge \zeta \}$ = {\bf dom} $V(\rho)$ which is
a convex set. If $\zeta \ge 0$, then $\theta(\rho)- \zeta w(\rho)$ is a concave
function and hence, $S_\zeta =\{\rho \in {\bf dom} V(\rho) | V(\rho)\ge \zeta \}=\{\rho \in {\bf dom} V(\rho) | \theta(\rho) - \zeta w(\rho)\ge 0 \}$ is a convex set since the superlevel sets of concave functions are convex.
\bibliographystyle{IEEEtran}
\bibliography{IEEEabrv,energy_bib}
\balance
\end{document}